\documentclass[12pt,preprint]{aastex}
\usepackage{emulateapj5}
\usepackage{apjfonts}
\usepackage{epsfig}
\usepackage{natbib}
\newcommand{\alp}{\ensuremath{\alpha}}

\newcommand{\feh}{[Fe/H]}
\newcommand{\feave}{\ensuremath{\langle {\rm Fe} \rangle}}

\def\gtrsim{\mathrel{\hbox{\rlap{\hbox{\lower4pt\hbox{$\sim$}}}\hbox{\raise2pt\hbox{$>$}}}}}

\newcommand{\hbeta}{H\ensuremath{\beta}}

\newcommand{\kms}{km~s\ensuremath{^{-1}}}

\newcommand{\mgb}{Mg$b$}

\newcommand{\msun}{\ensuremath{M_{\odot}}}

\newcommand{\oiii}{[\ion{O}{3}]}

\newcommand{\sers}{S{\'e}rsic}

\newcommand{\sigmastar}{\ensuremath{\sigma_{\ast}}}

\def\lax{{$\mathrel{\hbox{\rlap{\hbox{\lower4pt\hbox{$\sim$}}}\hbox{$<$}}}$}}
\def\gax{{$\mathrel{\hbox{\rlap{\hbox{\lower4pt\hbox{$\sim$}}}\hbox{$>$}}}$}}

\slugcomment{Accepted for publication in {\it The Astrophysical Journal}.}
\shorttitle{{\it MASSIVE II}}
\shortauthors{GREENE, ET AL.}

\begin{document}

\title{The MASSIVE Survey II: Stellar Population Trends Out to Large Radius in Massive Early Type Galaxies}

\author{Jenny E. Greene\altaffilmark{1}, Ryan Janish\altaffilmark{2}, 
Chung-Pei Ma\altaffilmark{3}, Nicholas J. McConnell\altaffilmark{4}, 
John P. Blakeslee\altaffilmark{5}, Jens Thomas\altaffilmark{6}, 
Jeremy D. Murphy\altaffilmark{7}}

\altaffiltext{1}{Department of Astrophysics, Princeton University, Princeton, NJ 08544, USA}
\altaffiltext{2}{Department of Physics, University of California, Berkeley, CA 94720, USA}
\altaffiltext{3}{Department of Astronomy, University of California, Berkeley, CA 94720, USA}
\altaffiltext{4}{Institute for Astronomy, University of Hawaii at Manoa, Honolulu, HI 96822, USA}
\altaffiltext{5}{Dominion Astrophysical Observatory, NRC Herzberg Institute of Astrophysics, Victoria, BC V9E 2E7, Canada}
\altaffiltext{6}{Max Planck-Institute for Extraterrestrial Physics, Giessenbachstr. 1, D-85741 Garching, Germany}
\altaffiltext{7}{IXL Learning, 777 Mariners Island Blvd., Suite 600, San Mateo, CA 94404}

\begin{abstract}
We examine stellar population gradients in $\sim 100$ massive early
type galaxies spanning $180 < \sigmastar < 370$~\kms\ and $M_K$ of
$-22.5$ to $-26.5$ mag, observed as part of the MASSIVE survey
\citep{maetal2014}. Using integral-field spectroscopy from the
Mitchell Spectrograph on the 2.7m telescope at McDonald Observatory,
we create stacked spectra as a function of radius for galaxies binned
by their stellar velocity dispersion, stellar mass, and group
richness. With excellent sampling at the highest stellar mass, we
examine radial trends in stellar population properties extending to
beyond twice the effective radius ($\sim 2.5 R_e$).  Specifically, 
we examine trends in age, metallicity, and abundance ratios of Mg, C, N,
and Ca, and discuss the implications for star formation histories and
elemental yields.  At a fixed physical radius of $3-6$
kpc (the likely size of the galaxy cores formed at high redshift)
stellar age and [\alp/Fe] increase with increasing \sigmastar\ and
depend only weakly on stellar mass, as we might expect if denser
galaxies form their central cores earlier and faster. If we instead
focus on $1-1.5 R_e$, the trends in abundance and
abundance ratio are washed out, as might be expected if the stars at
large radius were accreted by smaller galaxies.  Finally, we show that
when controlling for \sigmastar, there are only very subtle
differences in stellar population properties or gradients as a
function of group richness; even at large radius internal properties
matter more than environment in determining star formation history.

\end{abstract}


\section{Introduction}

The assembly history of elliptical galaxies remains a major unsolved
problem for galaxy evolution. Recent observations point to dramatic
size evolution of the most massive galaxies from $z \approx 2$ to the
present
\citep[e.g.,][]{vandokkumetal2008,vanderweletal2008,pateletal2013,
  vanderweletal2014}. The extent to which these trends require
late-stage minor mergers \citep[e.g.,][]{oseretal2012} or can be
explained by the addition of larger, younger galaxies at later times
\citep[][]{valentinuzzietal2010,newmanetal2012,barroetal2013} remains
a topic of ongoing debate. Information lurking in the faint outer
parts of present-day massive ellipticals can complement high-redshift
measurements.  Radial gradients in stellar populations distinguish
when and how the stars at large radius were formed
\citep[e.g.,][]{white1980,kobayashi2004,greeneetal2013,
  hirschmannetal2015}, while the kinematics of the stars (e.g.,
$V/\sigmastar$, the level of radial anisotropy, etc.) contain clues
about how these stars entered the halo
\citep[e.g.,][]{wuetal2014,arnoldetal2014,raskuttietal2014,
  naabetal2014,rottgersetal2014}.

Here we focus on the average radial trends in stellar populations of
early-type galaxies using our ambitious survey of the hundred most
MASSIVE galaxies within one hundred Mpc \citep{maetal2014}.
Observations of the stellar populations in elliptical galaxy outskirts
are challenging, since their surface brightnesses drop steeply with
radius.  Despite more than thirty years of effort, most observations
of stellar population gradients do not extend much beyond the
half-light radius
\citep{spinradtaylor1971,faberetal1977,gorgasetal1990,
  fisheretal1995,kobayashiarimoto1999,ogandoetal2005, broughetal2007,
  baesetal2007,annibalietal2007,sanchez-blazquezetal2007,rawleetal2008,
  kuntschneretal2010,mcdermidetal2015,olivaetal2015}. Resolved stellar population
studies have uncovered a low-metallicity halo component at very large
radius, but only in a handful of nearby galaxies
\citep[e.g.,][]{kaliraietal2006,harrisetal1999,rejkubaetal2005,
  harrisetal2007,crnojevicetal2013,pastorelloetal2014,peacocketal2015,
  williamsetal2015}.  There are also a few long-slit
observations that extend to large radius
\citep{carolloetal1993,carollodanziger1994,
  mehlertetal2003,kelsonetal2006,spolaoretal2010,puetal2010,puhan2011}.
Even integral-field spectrographs, now widely used for the study of
spatially resolved galaxy properties \citep{emsellemetal2004,
  sarzietal2006,cappellarietal2006,cappellarietal2012}, include few
observations that extend beyond the half-light radius in integrated
light \citep{weijmansetal2009,murphyetal2011}.

The $\sim 100$ galaxies analyzed here represent a significant
improvement over previous work.  Using coadded spectra as a function
of radius, we will investigate whether radial gradients depend not
only on stellar velocity dispersion, but also on other intrinsic
galaxy properties such as stellar mass and environmental density. We
find a hint that at fixed \sigmastar, galaxies residing in higher
densities are older and more \alp-element enhanced, but we see no
evidence for differing gradients for galaxies in low and high
densities. 

We present the sample in \S \ref{sec:Sample}, the observations and
data reduction in \S \ref{sec:Obs}, our stellar population analysis in
\S \ref{sec:Analysis}, and the radial variations in stellar
populations in \S \ref{sec:Radial}.  We summarize our findings and
conclude in \S \ref{sec:Summary}. Throughout we assume a concordance 
cosmology with $H_0=70$~km~s$^{-1}$~Mpc$^{-1}$, 
$\Omega_{\rm M}=0.3$ and $\Omega_{\rm \Lambda}=0.7$ 
\citep{dunkleyetal2009}.

\section{Sample}
\label{sec:Sample}

The MASSIVE sample selection is described in detail in
\citet{maetal2014}. For completeness, we summarize the sample
selection briefly here.  MASSIVE is a volume-limited sample of the 116
most massive galaxies within 108 Mpc. The galaxies are selected from
the 2MASS Redshift Survey \citep[2MRS;][]{huchraetal2012}, using a
total $K-$band magnitude limit of $M_K < -25.3$ mag (roughly
$10^{11.5}$~\msun).  Using the Hyperleda database
\citep{patureletal2003}, we apply a morphological cut to remove large
spiral and interacting galaxies. The resulting 116 galaxies span a
wide range of stellar velocity dispersion ($\sim 180-400$~\kms) and,
based on the group catalog of \citet{crooketal2007}, are found in a
wide array of environments, from ``field'' galaxies with no $\sim L^*$
companions to rich clusters (Coma, Perseus, and Virgo).  In this
paper, we include 49 MASSIVE galaxies with large-format integral-field
spectrograph observations in hand.  While we are also obtaining high
spatial resolution integral field spectroscopy for a subset of
galaxies for black hole mass determinations
\citep[e.g.,][]{mcconnelletal2012}, this paper deals exclusively with
the Mitchell data.

\begin{figure*}
\vbox{ 
\vskip -2mm
\hskip +35mm
\psfig{file=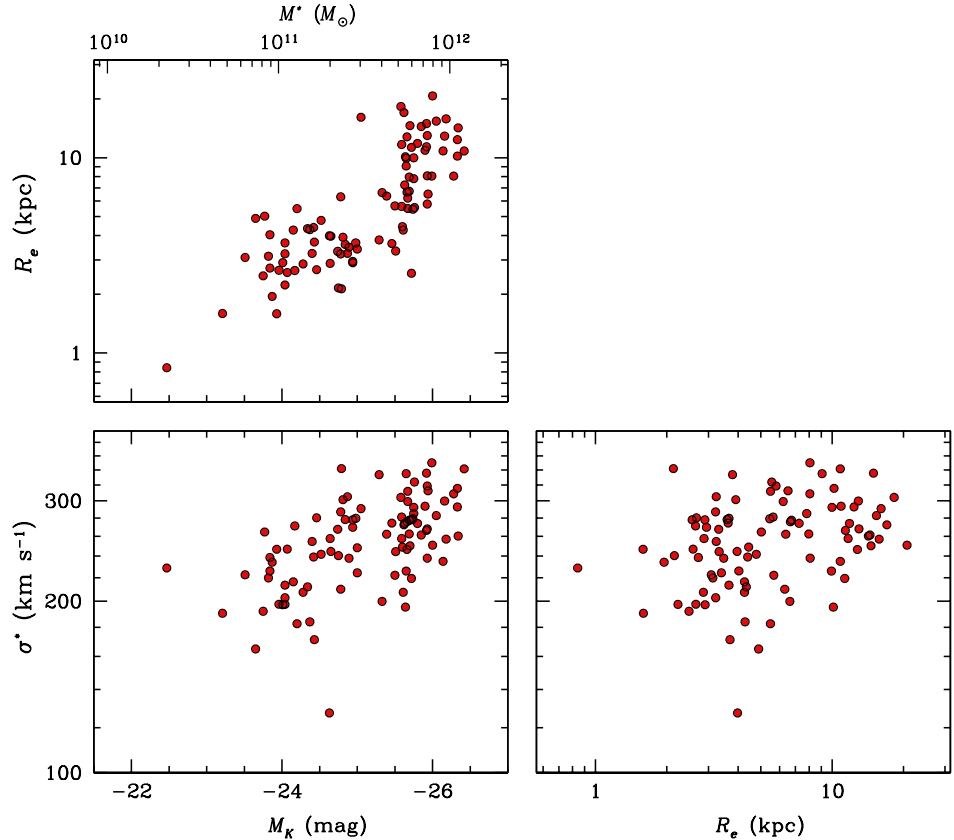,width=0.6\textwidth,keepaspectratio=true,angle=90}
}
\vskip -0mm
\figcaption[]{
Distributions of basic structural properties for our sample. On bottom
left, stellar velocity dispersion (\sigmastar) measured within the
central fiber is plotted against $M_K$ (mag; total
magnitude) from 2MASS. On top left, we show effective radius, from the
NSA catalog (or from 2MASS with a correction; \S 3.2) against $M_K$
and on bottom right effective radius versus stellar velocity
dispersion.}
\label{fig:FJ}
\end{figure*}

In addition to the ongoing MASSIVE survey, we include here a sample of
lower-mass galaxies \citep{greeneetal2013} selected directly from the
Sloan Digital Sky Survey \citep[SDSS;][]{yorketal2000} and observed
with an identical set-up with the Mitchell Spectrograph.  As the
selection was different, we briefly review it here \citep[see
  also][]{greeneetal2012,raskuttietal2014}. Since the spectral
resolution of the Mitchell Spectrograph is $\sigma_{\rm inst} \approx
150$~\kms\ at 4000\AA, we select galaxies with dispersion measurements
from the SDSS that are greater than this value. Individual fibers are
$4\farcs2$ in diameter, and so we aim for galaxies with effective radii
at least twice as large.  Galaxies with distances of 40-95 Mpc are
large enough to be well-resolved but small enough to fit into one
pointing.  We use a color selection of $u-r>2.2$
\citep[][]{stratevaetal2001}, which preferentially selects early-type
galaxies, and then remove the few edge-on disk galaxies by hand, but
keep S0 galaxies. These 46 galaxies are of uniformly
lower stellar mass than those in MASSIVE. Taken together, the sample 
galaxies span a
range of $-22.5 < M_K < -26.5$ Vega mag. The sample properties for 
the 95 galaxies (49 from MASSIVE) are summarized in Figure \ref{fig:FJ}.

\section{Observations and Data Reduction}
\label{sec:Obs}

The majority of the galaxies presented here were observed over ten
observing runs between Sept 2011 and April 2014.  There are an
additional four galaxies taken earlier as part of the PhD thesis of
Jeremy Murphy \citep{murphyetal2011} and four taken by Nicholas
McConnell \citep{mcconnelletal2012}. All of these data
were taken with the George and Cynthia Mitchell Spectrograph
\citep[the Mitchell Spectrograph, formerly VIRUS-P;][]{hilletal2008a}
on the 2.7m Harlan J. Smith telescope at McDonald Observatory. The
Mitchell Spectrograph is an integral-field spectrograph with 246
fibers subtending $4\farcs2$ each and covering a
107\arcsec$\times$107\arcsec\ field of view with a one-third filling
factor. As such, the Mitchell Spectrograph is ideal for studying the
low surface brightness outer parts of nearby galaxies
\citep{blancetal2009,yoachimetal2010,murphyetal2011,adamsetal2012}.

We utilize the blue setting of the Mitchell
spectrograph with a resolution of R $\approx$ 850 and spanning a
wavelength range of 3550-5850~\AA. This resolution (roughly a FWHM of
5\AA) delivers a dispersion of $\sim$~1.1~\AA\ pixel$^{-1}$ and
corresponds to \sigmastar~$\approx 150$~\kms\ at 4300~\AA, our bluest
Lick index, improving to $\sim 100$~\kms\ at the red end of the 
spectrum.  Each galaxy was observed for a total of $\sim 2$ hours on
source with one-third of the time spent at each of three dither
positions to fill the field of view. Observations are interleaved with
sky observations of 10 min duration. Initial data reduction is
accomplished using the custom code Vaccine
\citep{adamsetal2011,murphyetal2011}. The details of our data
reduction are described in \citet{murphyetal2011}, so we repeat only a
brief overview for completeness here.

We first perform overscan and bias subtraction for all science and
calibration frames. The fiber trace is determined from the twilight
flats, taking into account curvature in the spatial direction and
following the techniques of \citet{kelson2003} to avoid interpolation
and thus correlated errors.  Arcs are used to derive a wavelength
solution with typical rms residual variations about this best-fit
fourth-order polynomial between 0.05 and 0.1~\AA.  The twilight flats
are also used to construct the flat field, once the solar spectrum has
been modeled and removed. The flat field is typically stable to $<0.1$
pixels for typical thermal variations in the instrument.  The flat
field is then applied to all of the science frames to correct
variations in the pixel-to-pixel responses, as well as the relative
fiber-to-fiber variation, and the cross-dispersion profile shape for
every fiber.  The sky is modeled using off-galaxy sky frames observed
with a sky-object-object-sky pattern. The sky frames are processed in
the same manner as the science frames. In general, each sky nod is
weighted equally, except for very cloudy conditions. Finally, cosmic
rays are identified and masked.

We use software developed for the VENGA project
\citep{blancetal2009,blancetal2013} for flux calibration and final
processing.  We observe flux calibration stars using a six-point
dither pattern and derive a relative flux calibration in the standard
way. We can test the wavelength dependence of the flux calibration by
comparing the shape of the spectrum in the central fiber of the
Mitchell Spectrograph with the SDSS spectrum for those galaxies with
SDSS spectra.  We find $\lesssim 10\%$ disagreement in nearly all
cases, with no more than $\sim 15\%$ differences at worst.  We then
correct the default astrometric solutions using photometry, by
deriving an astrometric match between each image and an SDSS
\citep{yorketal2000} or PanSTARRS
\citep{schlaflyetal2012,tonryetal2012,magnieretal2013} image.
Finally, all fibers are interpolated onto the same wavelength 
scale and combined into radial bins.

\subsection{Effective radii}

In our previous work, we adopted the SDSS model radius (the de
Vaucouleurs fit) as the effective radius ($R_e$). While there is
considerable evidence that the shape of the light profile changes
systematically with galaxy mass
\citep[e.g.,][]{caonetal1993,ferrareseetal2006,kormendyetal2009},

\vbox{ 
\vskip -2mm
\hskip -2mm
\psfig{file=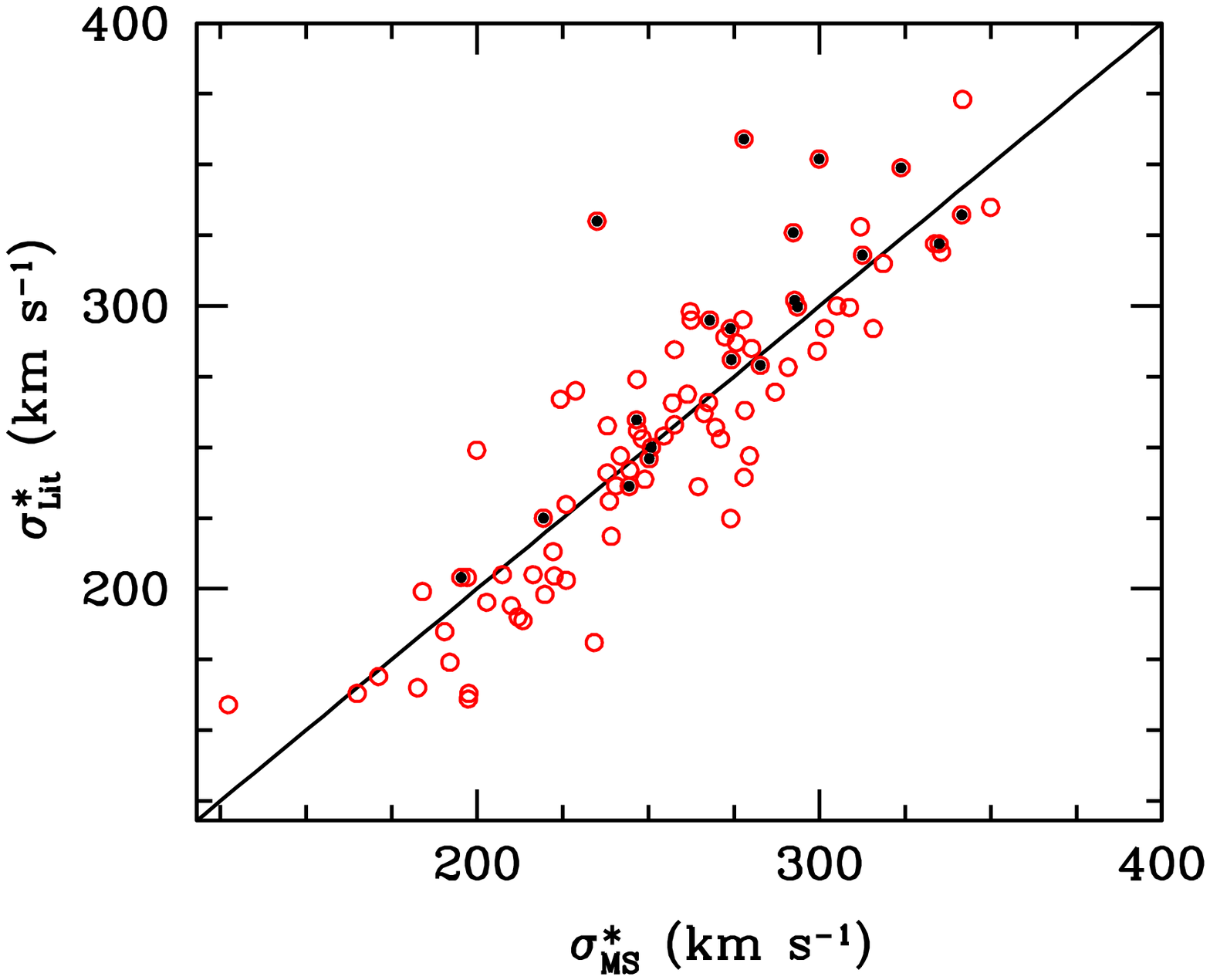,width=0.45\textwidth,keepaspectratio=true,angle=0}
}
\vskip -0mm
\figcaption[]{
Stellar velocity dispersions (\sigmastar) measured within 
the central $4\farcs2$ Mitchell fiber are compared with literature 
values. Open symbols have literature measurements from the SDSS, 
while filled symbols are from Hyperleda. The overall agreement is 
quite good, with (\sigmastar$_{\rm MS}-$\sigmastar$_{\rm
  Lit}$)/ \sigmastar$_{\rm Lit} = 0.01 \pm 0.09$.
\label{fig:sigvsig}}
\vskip 5mm

\noindent
fitting the galaxies with a fixed \sers\ index of four has the benefit
that we are less sensitive to both sky subtraction errors
\citep{mandelbaumetal2005,bernardietal2007} and to the detailed shape
of the light profile in the very faint wings
\citep[e.g.,][]{lacknergunn2012}.  In the case of MASSIVE, $\sim 1/3$
of the galaxies do not have SDSS imaging, so we adapt a size
measurement from 2MASS \citep{jarrettetal2003}.  We use the median
effective radius measured from the $JHK$ band \citep[see Eq. 3 in
][]{maetal2014}.  The 2MASS size measurement tends to underestimate
the galaxy sizes relative to the SDSS
\citep[][]{laueretal2007,kormendyetal2009}. To put the size
measurements on equal footing, \citet{maetal2014} fit a linear
conversion between $R_{\rm 2MASS}$ and $R_{\rm SDSS}: \rm{log}_{10}
(R_{\rm 2MASS}) = 0.8 \rm{log}_{10} (R_{\rm SDSS}) - 0.076$ (their
Eq. 4).  We use this relation to correct the 2MASS measurements to
match the SDSS measurements.

\subsection{Radial bins}

Spectra from individual fibers, with the exception of those at the
very center of the IFU, have inadequate signal for stellar population
studies.  Therefore, all of our analysis is performed on binned
spectra.  We utilize two binning schemes here, in both cases defining
elliptical annuli based on the axis ratio measured by the SDSS or
2MASS.  Since galaxies get larger as they get more massive, we make
bins of width $0.5 R_e$.  However, it is also interesting to look at
trends as a function of physical size, and here we extend from 0-15
kpc in 3 kpc increments. For the physical bins, we
use the central fiber as the central bin; while this central fiber
corresponds to a different physical size for each system, it provides
our highest spatial resolution bin.

\subsection{Stellar velocity dispersion measurements}

Stellar velocity dispersions are required for measuring Lick indices.
This is because at fixed intrinsic absorption, as the velocity
dispersion increases, the measured EW decreases.  Thus, a correction
must be applied to put all indices on the same scale. We also use the
central stellar velocity dispersions to rank galaxies in constructing
coadded spectra. 

\vbox{ 
\vskip +10mm
\hskip -2mm
\psfig{file=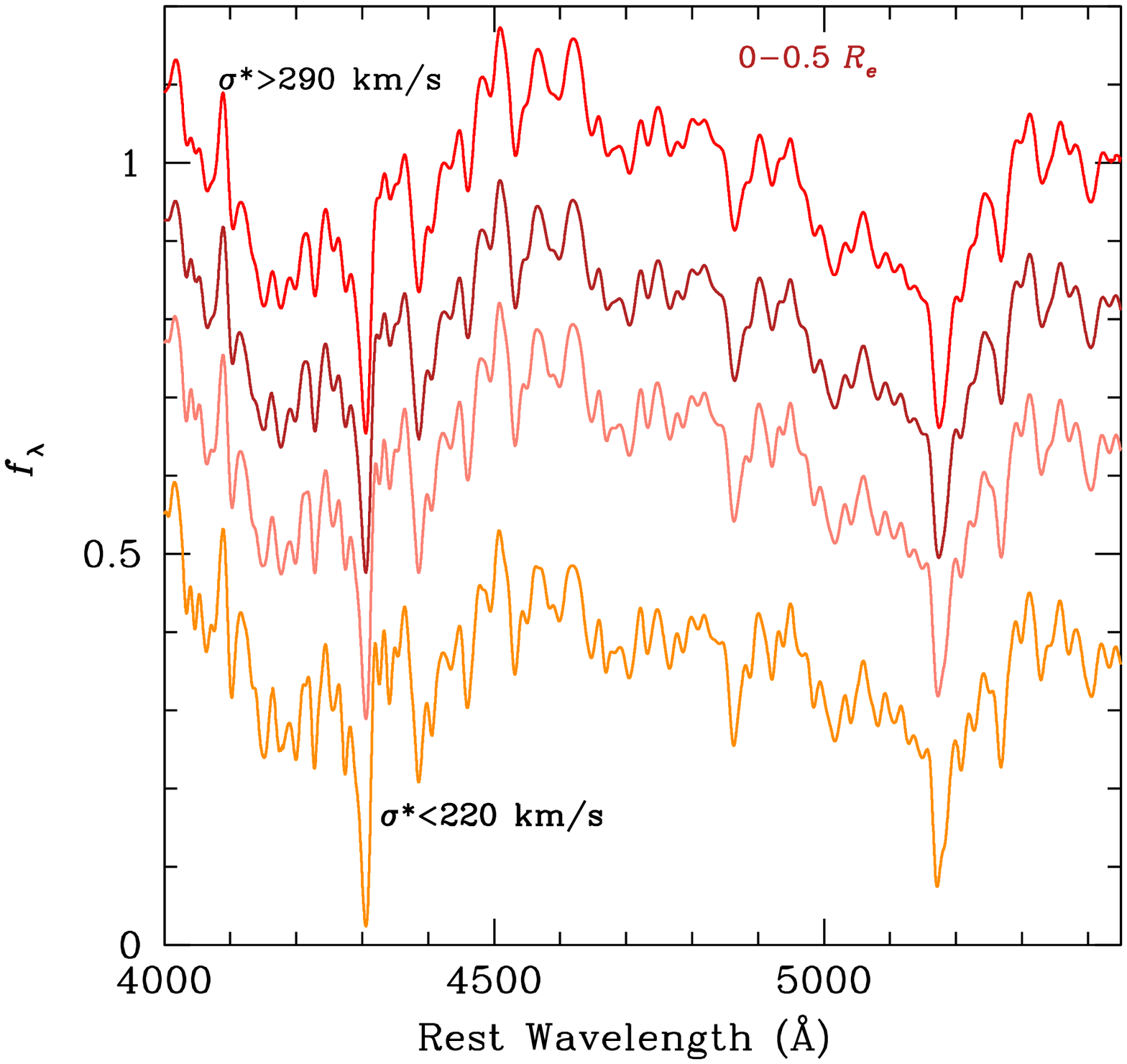,width=0.45\textwidth,keepaspectratio=true,angle=0}
}
\vskip -0mm
\figcaption[]{
The coadded spectra within $0.5 R_e$, for stellar velocity dispersion 
stacks starting from $\sigmastar < 220$~\kms\ (bottom) to 
$\sigmastar > 290$~\kms\ (top). Each spectrum has been normalized to 
unity as described in the text, and the flux density offsets are arbitrarily 
applied for display purposes. In creating the stacks, 
each galaxy was smoothed to a 
value 30\% more than the upper end of the dispersion range in 
creating the stack. 
\label{fig:specstack}}
\vskip 5mm

\noindent
Our central $4\farcs2$ fiber is similar to the SDSS
fiber and thus easily compared with the literature. We use pPXF to
measure the dispersions \citep{cappellariemsellem2004}.  We compare
our \sigmastar\ measurements with SDSS measurements when available or
our compiled literature values \citep{maetal2014} derived from the
Hyperleda database \citep{patureletal2003}. We find very good
agreement, with the median ($\sigma_{\rm MS}-\sigma_{\rm
  Lit}$)/$\sigma_{\rm Lit} = 0.01 \pm 0.09$.  That is, we see no
systematic offset between the two sets of measurements and a scatter
of 9\%. We compare our measurements with the literature in Figure
\ref{fig:sigvsig}.  As discussed in \citet{vandenboschetal2015}, there
is some excess scatter at the high dispersion end, specifically when
we compare with Hyperleda. However, we find better agreement with
Hyperleda than reported by van den Bosch et al.\ predominantly due to
recent (post-2013) changes in Hyperleda.  We have 51 objects in common
with the van den Bosch HET catalog, and we also find reasonable
agreement with their \sigmastar\ measurements, with ($\sigma_{\rm
  MS}-\sigma_{\rm HET}$)/ $\sigma_{\rm HET} = -0.03 \pm
0.07$. Throughout, \sigmastar\ refers to our measurements from the
central fiber unless otherwise specified.

\section{Analysis}
\label{sec:Analysis}

\subsection{Stellar population modeling approach}

We use Lick indices as a tool to trace the stellar populations.  Lick
indices were developed as a way to extract stellar population
information from spectra without flux calibration, but still
circumvent classic age-metallicity degeneracies
\citep{burstein1985,faberetal1985,wortheyetal1992,trageretal1998}.  The
Lick indices are narrow regions of the spectra (typically $\sim
20$\AA\ wide) that are dominated by a single element and thus are
predominantly sensitive to orthogonal aspects of the stellar
population properties; \hbeta\ is sensitive to age, Fe indices to
[Fe/H], and so on.  It is important to bear in mind that 
at the velocity dispersion of our target
galaxies (200-400 \kms) all indices are blends of multiple
elements. See, for example, Table 1 in \citet{gravesschiavon2008} for
the primary elements that dominate the Lick indices used in this work.

Lick indices are still widely used in the literature as they mitigate
difficulties in modeling the effects of abundance ratio changes
\citep[e.g.,][]{wortheyetal1994,gallazzietal2005}.
\citet{trageretal2000a,trageretal2000b} developed a technique to
derive index responses from stellar atmosphere models, such that even
if the full spectrum cannot be calculated, the index EW changes due to
changing abundance ratios can be incorporated into an analysis of line
EWs. On the other hand, the current generation of full spectral
synthesis codes are very sophisticated
\citep[e.g.,][]{vazdekisetal2010,conroyvandokkum2012mod} and have been
designed to fit non-solar abundance ratios.

For reference, we review the main indices that we use to derive the
basic stellar population parameters. The reader is referred to
\citet{gravesschiavon2008} for more detailed information.  We use {\it
  lick\_ew} \citep{gravesschiavon2008} to measure the Lick indices and
the stellar population modeling code {\it EZ\_Ages}
\citep{gravesschiavon2008} to convert the Lick indices to physical
parameters (age, \feh, [\alp/Fe]). The code works on a hierarchy of
index pairs, starting with \hbeta\ and \feave, and iteratively solves
for the age, abundance and abundance ratios.  The models of
\citet{schiavon2007} include abundance ratio differences using 
the response functions of
\citet{kornetal2005}.  For a different inversion methodology see
\citet{thomasetal2011} or for full spectral fitting comparisons see
\citet{conroyetal2014}.

As emphasized by \citet{schiavon2007}, because we do not directly
measure the oxygen abundance and oxygen is the most abundant heavy
element, it is misleading to quote total metallicity. Instead, we
quote [Fe/H], which is directly inferred from the Fe indices.  If we
assume that [O/Fe] tracks [Mg/Fe], then we can use the latter to infer
[Z/H].  We will generally assume that O and Mg follow similar trends
as they are both \alp\ elements, and thus use [Mg/Fe] interchangeably
with [\alp/Fe].  Based on the same assumption, we also will use the
conversion from \citet{trageretal2000a}: [Z/H]=[Fe/H]+0.94[\alp/Fe] to
calculate the metallicity.  For alternate approaches to modeling
oxygen using Lick indices, see \citet{johanssonetal2012} or
\citet{wortheyetal2014}. In our default runs, we utilize the
\alp-enhanced isochrone from \citet{salasnichetal2000} and the default
assumption that [O/Fe]$=0.5$ to match the \alp-enhanced isochrone
value.

\begin{figure*}
\vbox{ 
\vskip -9mm
\hskip +10mm
\psfig{file=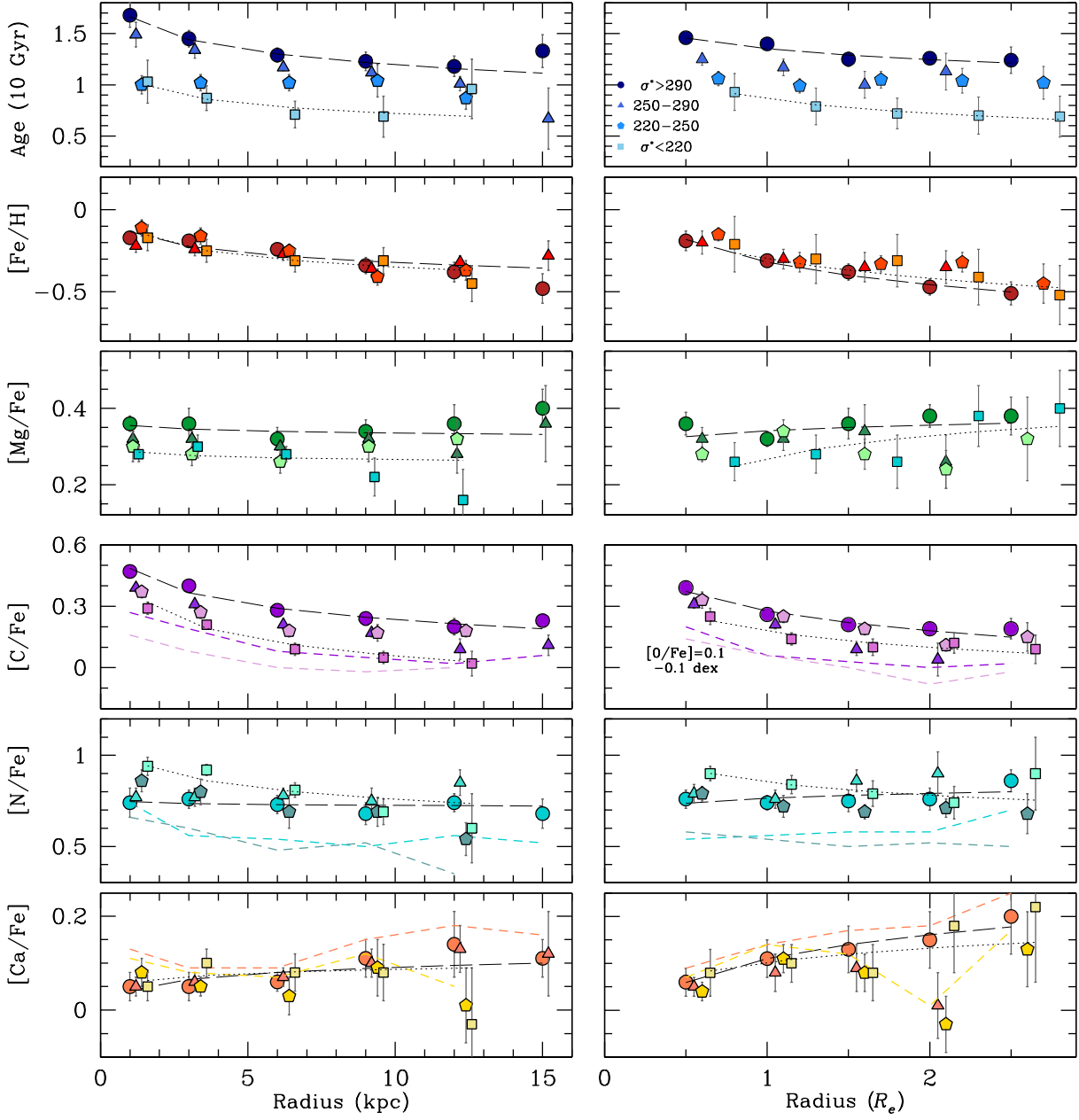,width=0.9\textwidth,keepaspectratio=true,angle=-90}
}
\vskip -0mm
\figcaption[]{
Radial gradients in age, \feh, [Mg/Fe], [C/Fe], [N/Fe], and [Ca/Fe] as
calculated by \emph{EZ\_Ages} from the Lick indices measured in the
coadded spectra.  The measurements are made on four stacked spectra
binned on stellar velocity dispersion (see figure key), and are shown
as a function of $R$ in kpc (left) or $R/R_e$. 
We fit the radial gradients with a power law of the form 
$X = {\rm A~log} (R/R_3) + {\rm B}$ for each stellar population 
parameter $X$, where $R_3$ is either $3-6$ kpc or $1-1.5 R_e$. 
The fits to the highest (long-dashed lines) and 
lowest dispersion (dotted lines) are shown here, and in 
Tables 1 \& 2. Note
the decline with radius in \feh\ and [C/Fe] in contrast with the
radially constant age, [Mg/Fe], [N/Fe], and [Ca/Fe]. 
To indicate systematic errors in the light elements due to the 
unknown oxygen abundance, we also show the resulting models
assuming [O/Fe]$=0.1$ rather than the default
[O/Fe]$=0.5$ (keeping [O/Fe] constant with radius in both cases;
$\sigmastar> 290$~\kms\ coadd in dark and $220 < \sigmastar < 250$
~\kms\ in light color). The [C/Fe] lines with alternate oxygen abundance 
have been offset by $-0.1$ dex for presentation purposes.
\label{fig:radialagefeafe}}
\end{figure*}

Carbon is roughly half as abundant by number as oxygen at solar
abundances \citep{asplundetal2009}. We derive [C/Fe] from the
C$_24668$ index
\citep[e.g.,][]{tripiccobell1995,trageretal1998,gravesetal2007,
  johanssonetal2012}. Because the amount of C locked into CO depends
on the oxygen abundance, as we lower the assumed O abundance, the
inferred C abundance drops commensurately \citep[][]{servenetal2005}.
We will quantify the magnitude of this effect below.

Nitrogen (one-tenth the O abundance by number in the Sun) is then
derived from CN1 ($4143- 4178$\AA) once we have a C abundance. As
discussed in \citet{greeneetal2013}, there are some uncertainties
associated with CN due to the low S/N at the blue end of the spectrum,
but our nuclear CN measurements match those from the SDSS spectra of
the same galaxies (our Mitchell indices are $0.01 \pm 0.02$ mag lower 
than the SDSS measurements), giving us some confidence in these 
measurements. 

Finally, the Ca abundance (one-hundredth the C abundance by number in
the Sun) is based on the Ca4227 index \citep[$4223.5- 4236.0$\AA;
][]{worthey1998}.  This index is blended with CN, and thus the [Ca/Fe]
measurement is most uncertain, as it is dependent on both the [C/Fe]
and [N/Fe] measurements. Others have used Ca H+K
\citep{servenetal2005,wortheyetal2014} or the Ca Triplet at 8600\AA,
where the last is also sensitive to the dwarf-to-giant ratio
\citep[e.g.,][]{cenarroetal2004}.
 
Since the Lick indices are very sensitive to small errors in sky
subtraction and other small-scale errors, we construct stacked spectra
and measure the average radial trends in the Lick indices and
resulting stellar populations. 

\subsection{Equivalent widths and emission line corrections}

There are a number of systematic effects that may impact the
equivalent widths (EW). One relates to the velocity dispersion: while the
default dispersion corrections within \textit{lick\_ew} were verified
originally over the range $\sigma_* = 250-300$~\kms, some of our
galaxies exhibit even larger $\sigma_*$.  We therefore check our
dispersion corrections using simple stellar population models from
\citep{conroyvandokkum2012mod}, broadened over the full observed range of
$\sigma_*$.  The Lick indices that we recover using the default
\textit{lick\_ew} corrections agree with the input values within
0.02\AA\ for an \alp-enhanced model with a Salpeter IMF, which
is a good approximation to our galaxies. The corrections are a weak
function of stellar population parameters, but only at the hundredths
of an \AA\ level, which is small compared to our other sources of
systematic error.

A larger correction must be made for low-level emission that can fill
in the absorption lines and artificially lower their equivalent widths
(EWs). Weak emission from warm ionized gas is very common in the
centers of elliptical galaxies \citep{sarzietal2010,yanblanton2012},
and small amounts of line infill can lead to significant errors in
recovered parameters.  Even 0.1~\AA\ errors in \hbeta\ EW can lead to
errors of $\sim 1-2$ Gyr in the modeling \citep[e.g.,][]{schiavon2007}.

Given the very low emission levels, and the large uncertainties
involved, we compare two methods for determining the levels of
\oiii\ and \hbeta\ emission.  In \citet{greeneetal2012}, we utilized
pPXF+GANDALF developed by M. Sarzi \citep{sarzietal2006} and
M. Cappellari \citep{cappellariemsellem2004} to simultaneously model
the stellar absorption and emission lines. GANDALF is very robust
for well-detected lines, but under-constrained for very weak
emission. Following \citet{greeneetal2013}, we also fit each spectrum
with an empirical template drawn from the composite spectra of
\citet{gravesetal2010}.  We then fit the \oiii\ emission in the
residual spectrum, and subtract both \oiii\ and \hbeta, assuming that
the \hbeta\ emission is 70\% of the \oiii\ flux \citep[measured to
  within a factor of two,][]{trageretal2000a,gravesetal2007}. We then 
iterate these fits until the emission line flux has converged. In
addition, we search for residuals around strong sky lines at 5200 and
5460 \AA.

From our iterative fits, and focused on the galaxy centers for
simplicity, roughly two-thirds of the galaxies have low-level
\hbeta\ emission detected, with a median EW of 0.2\AA, and a maximum
of 1\AA\ (calculated for those galaxies with detected \hbeta).  
The Gandalf measurements do not
correlate very strongly with our iterative fits.  In the GANDALF fits,
only half of the galaxies have detections, with a median EW is
0.2\AA\ and a maximum \hbeta\ EW of 1.7\AA. As a means of
quantifying our systematic errors, we rerun the stacking analysis
described below on the Gandalf-subtracted spectra. As expected, only
the stellar age changes significantly, being $\sim 3$ Gyr lower in the
Gandalf stacks.  But all other stellar population properties are
virtually identical (with the \feh\ shifting higher by a small amount
to compensate the shift in stellar age).  Thus, we present results
based on the iterative fits, but we caution that there is a rather
large systematic uncertainty in the absolute stellar age.  All other
stellar population properties are robust to this modeling difficulty,
and in general the relative ages are robust as well. We are currently
working on more robust gas detection schemes using all lines in the
spectra (V. Pandya et al.  in preparation).

We then use {\it lick\_ew} \citep{gravesschiavon2008} on the
emission-line corrected spectra. The indices are on a modified
Lick system presented by \citet{schiavon2007} based on flux-calibrated
spectra.  In order to demonstrate that we are on the same system, we
compare the Lick indices from the flux-calibrated SDSS spectra (the
inner 3\arcsec) with those from the central $4\farcs2$ fiber in our
data. There is no net offset between the two sets of indices in any
case, with $\langle ({\rm H\beta \, EW}_{\rm S} - {\rm H \beta \,
  EW}_{\rm MS})/ {\rm H\beta \, EW}_{\rm MS} \rangle = 0.08 \pm 0.17$,
where S is SDSS and MS is the Mitchell Spectrograph.
$\langle$Fe$\rangle$ and \mgb\ each have a scatter of only 
$\sim 10\%$ and even smaller net offsets.

\subsection{Composite spectra}
\label{sec:Composite}

While measuring Lick indices is a very powerful technique for high S/N
spectra, at the large radii that we are working, systematic effects
such as small errors in sky subtraction and flux calibration can begin
to cause large uncertainties in the Lick indices measured from
individual objects. Stacked spectra
average over sky subtraction and flux calibration errors in individual
systems, which occur at different wavelengths in each galaxy
rest-frame \citep[e.g.,][]{gravesetal2009,yan2011}. Of course, 
variations in stellar populations at a given 
\sigmastar\ or mass are expected based on differences in accretion 
history \citep[e.g.,][]{hirschmannetal2015} and we are quite 
interested in these differences, particularly as a function of 
the dynamical properties of the galaxies. We plan to implement 
full spectral fitting in the
near future, which is more robust at low signal-to-noise 
ratio \citep[e.g.,][]{choietal2014}. 

We know that stellar population properties are a strong function of
\sigmastar\ \citep[e.g.,][]{wortheyetal1992,benderetal1993,
  trageretal2000b,gravesetal2009}.  Thus, we first divide the galaxies
into four stellar velocity dispersion bins using our pPXF measurements
to the central fiber.  Of the 95 galaxies in our sample, we exclude
eight. Most of these are from the low-mass sample and have a bright
star in the foreground making it difficult to reach a reasonable
S/N. NGC1167 in the MASSIVE sample is excluded because it may be a
face-on disk galaxy. We will treat it more carefully in future work.
The bins have \sigmastar$<220$~\kms\ (15 objects), $220
<$\sigmastar$<250$~\kms\ (21 objects), $250 <
$\sigmastar$<290$~\kms\ (33 objects), and all those higher than
\sigmastar$ > 290$~\kms\ (18 objects). Stacked spectra within $0.5
R_e$ are shown in Figure \ref{fig:specstack}. The resulting stellar
population parameters are shown in Figure \ref{fig:radialagefeafe}. We
reach different physical radii for different bins as we run out of
signal for the smaller and lower-mass galaxies. Also, the total number
of galaxies included in the largest radial bin is typically $\sim
30\%$ smaller than the centers, as various systematic effects such as
foreground star contamination grow more severe at low flux levels.

Below, we will also bin on stellar mass and group richness.  While the
detailed bins are different, the stacking technique described here is
the same for these different sets of bins.

To create the stack, we coadd the emission-line--subtracted spectra.
We interpolate the rest-frame spectra onto a common wavelength grid.
To ensure we know the dispersion of the final stacked spectrum, we
then smooth each galaxy to a value that is 30\% higher than the upper
\sigmastar\ limit of each bin. This smoothing ensures that all
galaxies go into the stack with the same effective dispersion. With
this approach, we minimize small errors due to dispersion corrections
to the Lick indices.  An alternate procedure would be to smooth all
templates, in all bins, to a high dispersion (e.g., 400~\kms). We find
very small ($<0.01$ \AA) differences in the resulting indices if we
adopt the latter approach, again confirming that our dispersion
corrections are working (\S 4.2). When we make bins in $M_K$ below, 
because of the wide range of \sigmastar\ in each bin, we smooth all 
bins to 400~\kms.

We remove the continuum by dividing each spectrum by a heavily
smoothed version of itself. This step simultaneously normalizes all
spectra to the same level and ensures that differences in continuum
shape (whether real or due to small errors in sky subtraction or flux
calibration) do not impact the final line strengths.  We then
calculate the median flux at each pixel, with rejection, although we
get very similar results using the biweight estimator
\citep[][]{beersetal1990}.  We experiment with multiplying the coadded
spectrum by the median continuum before measuring indices, but the
changes to the Lick indices are negligible.

\subsection{Uncertainties}

To determine the level of variation in the composite spectra, we
generate 100 boot-strapped composite spectra by randomly drawing from
the total list of galaxies in that bin, with replacement. We measure
Lick indices from each of these 100 trial spectra. We then assign
errors on the Lick indices that enclose 68\% of the Lick indices
measured from the 100 trials. Therefore, the size of the error is most
directly related to the variance in parameters over the population in
that bin, rather than measurement or modeling uncertainty. The
systematic uncertainties are not shown. Age measurements have large
systematic uncertainties due to emission infill (\S 4.2), while
nitrogen and calcium are particularly uncertain due to their
dependence on blue spectral features and assumed carbon and oxygen
abundances.

In general, we report the measurement from the primary stacked
spectrum, and the errors derived from the boot-strapped spectra.
There are a few cases where the Lick index measured from the primary
stacked spectrum does not fall within the stellar population grids
(usually because the \hbeta\ index is slightly too low). In these
cases, we use the median index value from the 100 trials as the final
answer. These cases are indicated with open symbols.

\begin{figure*}
\vbox{ 
\vskip -4mm
\hskip +15mm
\psfig{file=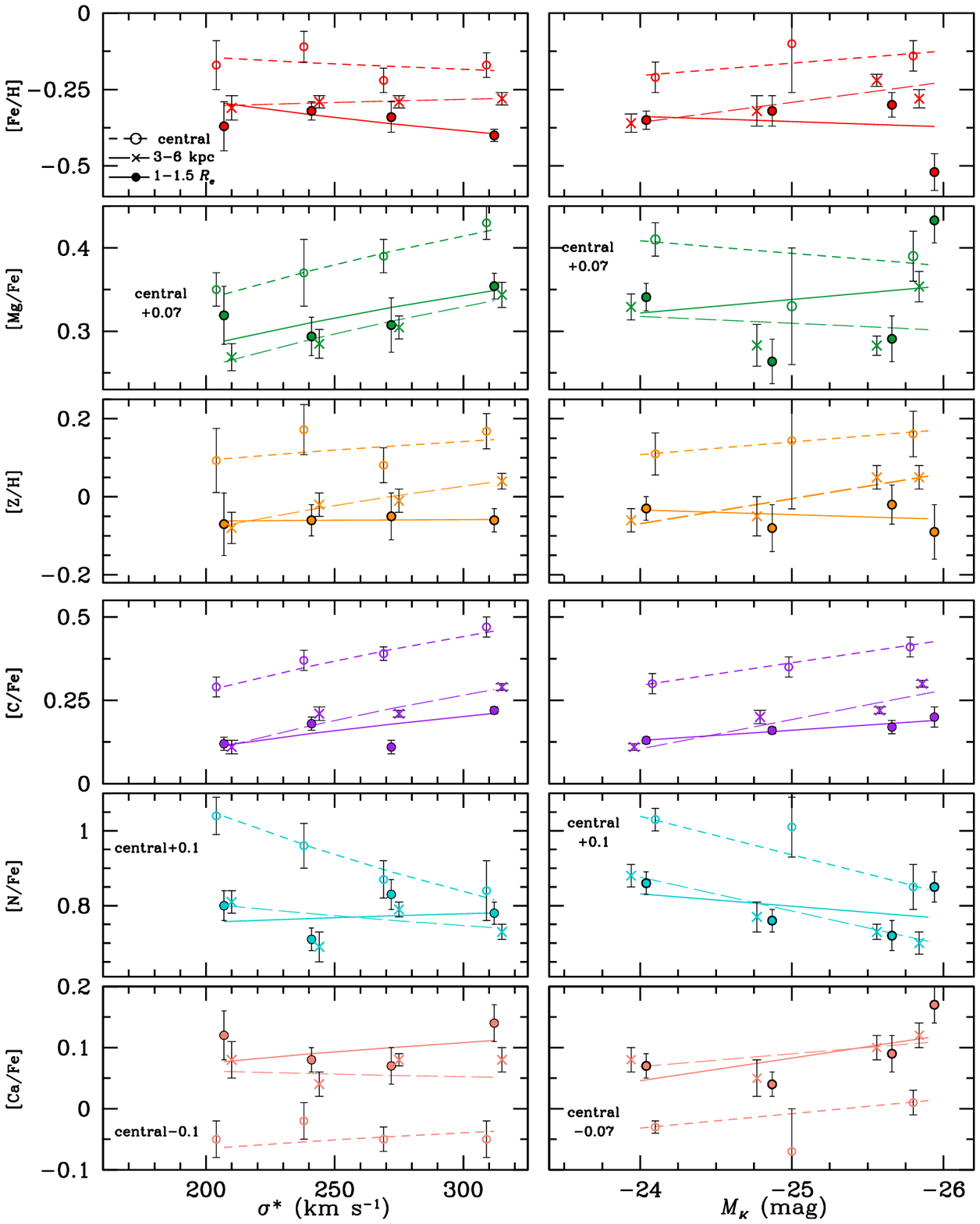,width=0.81\textwidth,keepaspectratio=true,angle=0}
}
\vskip -0mm
\figcaption[]{
{\it Left}: We compare the relation 
measured from the coadded central fibers (open circles; dashed lines) 
with those from fits to the profiles weighted to $1-1.5 R_e$ 
(filled circles; solid lines) and $3-6$ kpc (crosses; long-dashed 
lines).  We plot the fitted zeropoints (B) for each chemical
property fitted versus radius (Tables 1 \& 2) as a function of
\sigmastar. [Z/H] is estimated as 
[Z/H]=[Fe/H]+0.94[\alp/Fe], which assumes that
[O/Fe] tracks [Mg/Fe]. The power-law fits shown here have 
the form $X = {\rm A~log~(} {\sigma^{\ast}/250 {\rm \, km~s^{-1}})} + {\rm B}$ 
for each stellar population property $X$ (Table 3). 
We see that the central and $3-6$ kpc 
fits are generally similar and show the strongest trends with 
\sigmastar, while when we scale to $R_e$ the trends are washed 
out, supporting a scenario where the inner 5 kpc collapsed quickly 
and scales with the galaxy potential, while the outer parts are 
accreted from lower-mass systems.
{\it Right}: As above, but fitting 
$X = {\rm A~log~(} M_K/-25.3~{\rm mag)} + {\rm B}$. Note that 
the central measurement is missing for the highest $M_K$ bin because 
the \hbeta\ measurement fell off the grid.
In general the dispersion in galaxy properties is larger in fixed 
$M_K$ bins, and the trends weaker. The largest exception is [C/Fe] 
which shows a stronger correlation with $M_K$ than $\sigma^{\ast}$.
\label{fig:sigmasscor}}
\end{figure*}

\section{Radial Variations in Stellar Populations}
\label{sec:Radial}

\subsection{Bins of \sigmastar}
\label{sec:coadd_stellarpop}

In Figure \ref{fig:radialagefeafe} we show radial trends in the
measured age, \feh, and abundance ratios as a function of physical
(left) and $R_e$-scaled (right) radii.  Our default models make the
assumption that [O/Fe] is enhanced like [Mg/Fe], since they are
both \alp\ elements. However, in Figure \ref{fig:radialagefeafe} we
indicate with dashed lines the C, N, and Ca abundances that result for
an assumed solar [O/Fe].  While the zeropoints of [C/Fe] and [N/Fe]
both decline, the radial trends will not change unless [O/Fe]
(unlike [Mg/Fe]) changes with radius.

We fit a power-law relation between radius and each stellar population
property.  We anchor the relation at the center of our radial
coverage.  Eventually, when we have reliable $K-$band light profiles
for each galaxy, we will calculate mass-weighted stellar population
properties, but at present, for a given stellar population property
$X$ we fit a log-linear relationship : $X = {\rm A~log} (R/R_3) + {\rm
  B}$, where $R_3$ is the third bin, corresponding either 
to 3-6 kpc or $1-1.5 R_e$. The quantity A
represents the radial gradient per log radius. The quantity B
represents an effective stellar population property at the center of
our radial coverage.  The fits to $A$ and $B$ for each relation are
shown in Tables 1 \& 2.

The strongest radial gradients are found between \feh\ and [C/Fe],
which both decline with radius.  Age shows a decline in the highest
dispersion bin, but because that effect is weaker when we use the
alternate emission line correction, we treat age gradients with extra
caution. All other abundance ratios measured here are consistent with
remaining flat over the full radial range, aside from a $2\sigma$
increase in [Ca/Fe] at large radius that is only seen 
in the largest \sigmastar\ bin.

We turn to trends between stellar populations and \sigmastar. 
Our central bins are shown as open circles
in Figure \ref{fig:sigmasscor}.  We recover well-known trends
between stellar population properties and stellar velocity dispersion
for galaxy centers. Galaxies with higher stellar velocity
dispersions have older stellar ages and higher [Mg/Fe] and [C/Fe]
ratios \citep[e.g.,][]{trageretal2000a,worthey2004,thomasetal2005,
  sanchezblazquezetal2006,gravesetal2007,
  smithetal2009,priceetal2011,johanssonetal2012,
  wortheyetal2014,conroyetal2014}. One interesting exception is
[N/Fe], where we see a decline with \sigmastar.  This trend is at odds
with our previous finding \citep{greeneetal2013}, as well as most
other work on the topic \citep[although see
  also][]{kelsonetal2006}. More detailed work is needed to confirm
this trend, particularly given possible uncertainties in flux
calibration in the blue for the Mitchell spectra.
Recall also that the very old ages found
in the very centers of these galaxies have 2-3 Gyr error bars, and
thus are consistent with the age of the universe. 

Thanks to our spatial coverage, we can go beyond the trends between
\sigmastar\ and stellar populations in the galaxy center, and look at
how these trends evolve as we look to larger radius, both in physical
and $R_e-$scaled units. The crosses in Figure \ref{fig:sigmasscor}
show the trends between \sigmastar\ and stellar population properties
measured at $3-6$ kpc. In general, the same trends are seen with
\sigmastar\ in the central and $3-6$ kpc bins ([N/Fe] is again an
exception). Fit values are included in Table 3.  [C/Fe] shows the most
significant correlation with \sigmastar, with a positive slope
detected at $8 \, \sigma$ significance.  We find a weak trend with
[\alp/Fe] ($3.5\,\sigma$) and a strong trend with age ($7 \, \sigma$;
not shown).  The measured slopes between \sigmastar\ and \feh,
[Ca/Fe], and [N/Fe] are all consistent with zero; that is, the
effective value of these parameters is similar for all
\sigmastar\ bins.

We can also evaluate trends between \sigmastar\ and 
stellar population properties evaluated beyond $R_e$ (taken 
here to be the $1-1.5 R_e$ bin). In this case, the physical radii 
increase for the higher-dispersion bins. Interestingly, 
we find that the trends with \sigmastar\ are weaker 
when evaluated as a function of $R_e$. 
Only stellar age ($7 \,\sigma$) and [C/Fe] (much more weakly at
$4 \,\sigma$) correlate positively with \sigmastar. Apparently, trends
between \sigmastar\ and abundance ratios are strongest when measured
within small physical radii. If galaxies indeed form a compact core 
rapidly at high redshift, then we expect this inner region 
\citep[$<5$ kpc; e.g.,][]{vanderweletal2014} to depend most strongly 
on \sigmastar.

Before we interpret these observed trends in more
detail, we examine differences between bins in \sigmastar\ and stellar
mass.

\begin{figure*}
\vbox{ 
\vskip -1mm
\hskip +5mm
\psfig{file=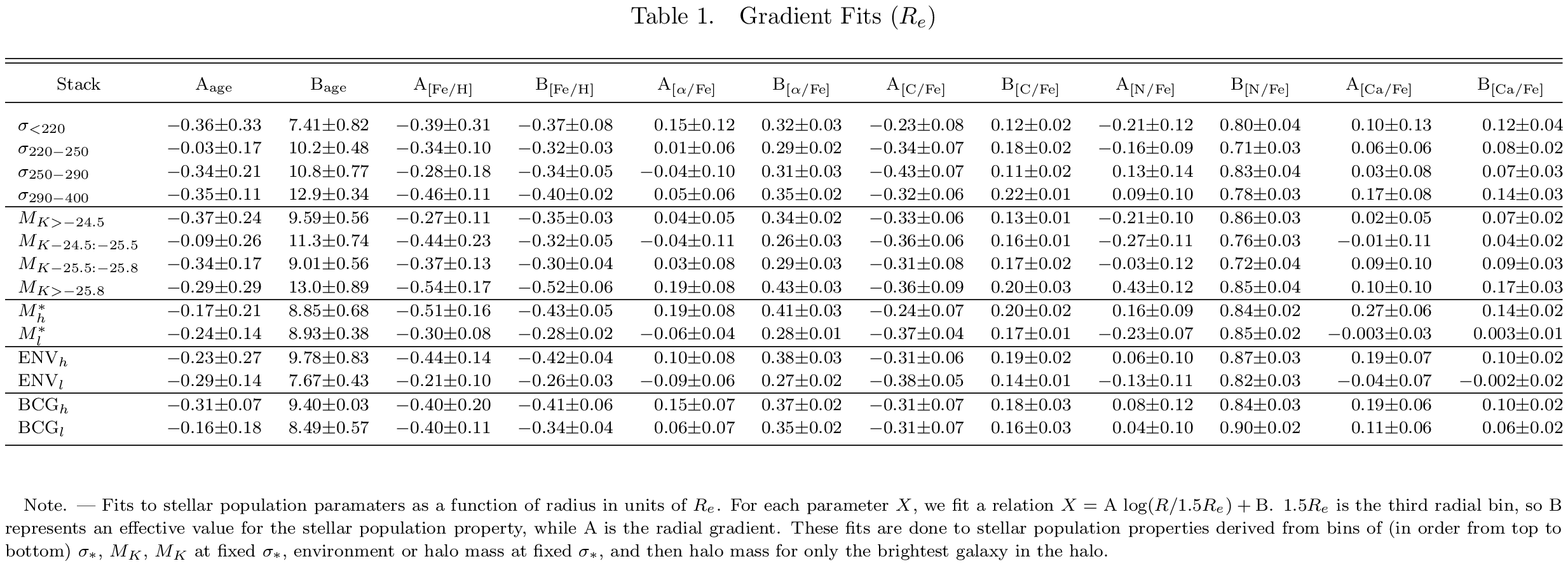,width=0.35\textwidth,keepaspectratio=true,angle=-90}
}
\vskip -0mm
\end{figure*}
\vskip 5mm

\begin{figure*}
\vbox{ 
\vskip -5mm
\hskip +5mm
\psfig{file=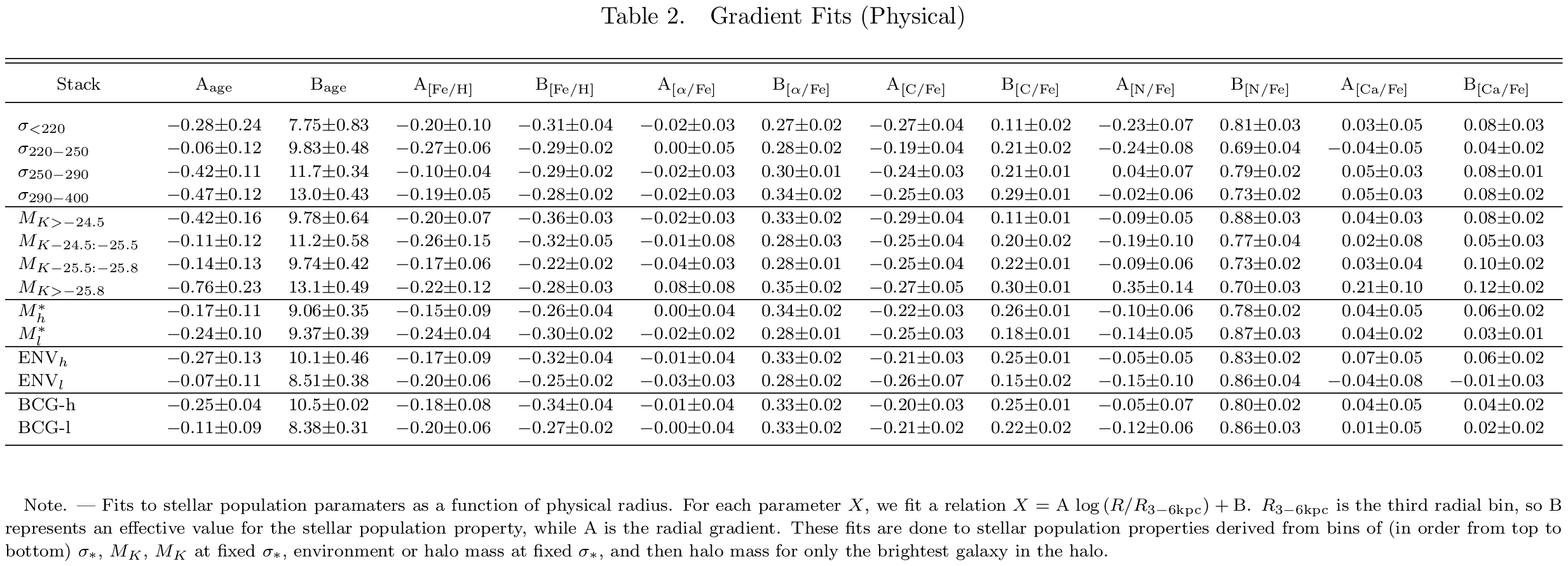,width=0.32\textwidth,keepaspectratio=true,angle=-90}
}
\vskip +2mm
\end{figure*}

\subsection{Bins of $M^*$}

\begin{figure*}
\vbox{ 
\vskip -5mm
\hskip +10mm
\psfig{file=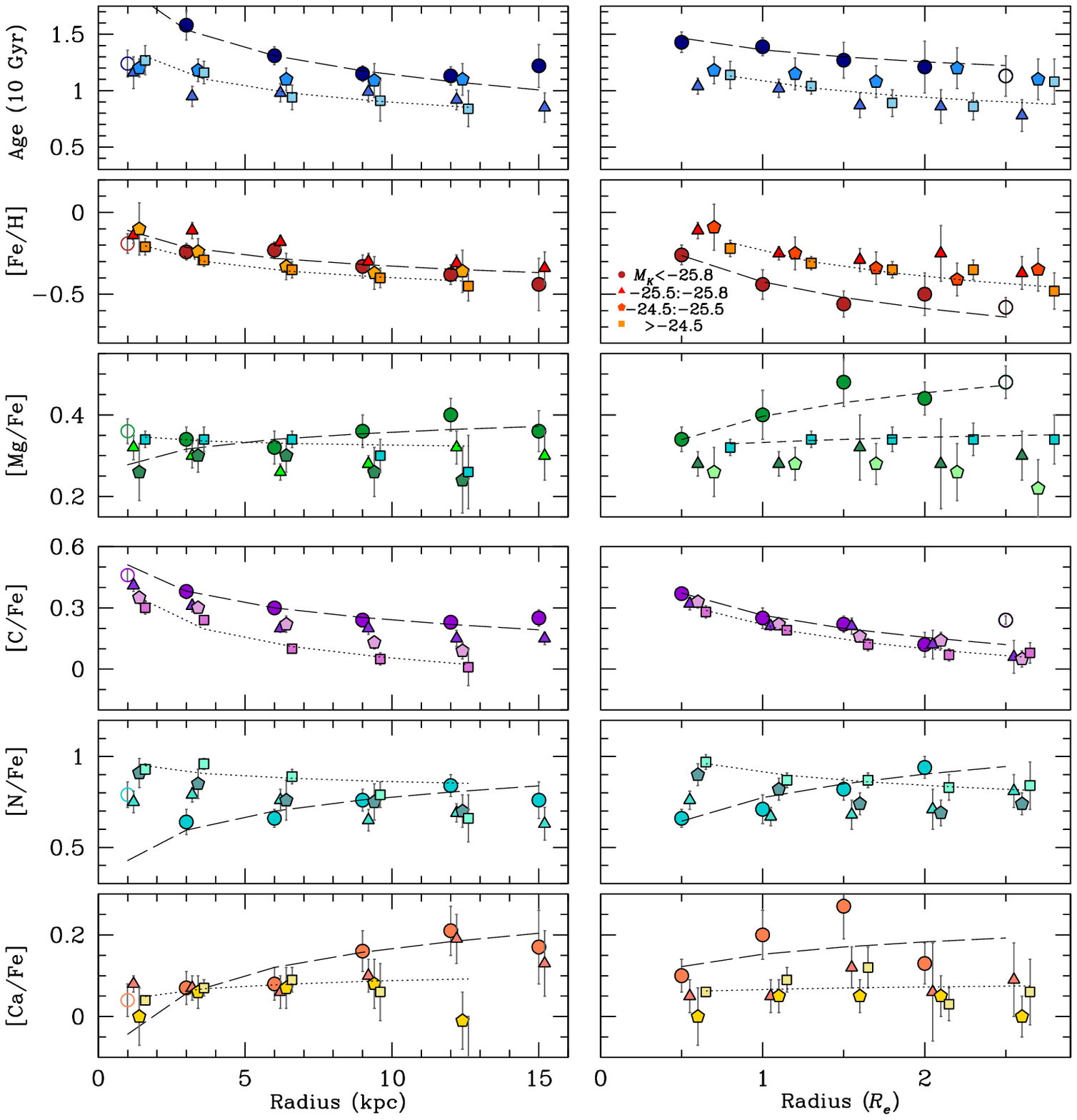,width=0.8\textwidth,keepaspectratio=true,angle=-90}
}
\vskip -0mm
\figcaption[]{
Radial gradients in age, \feh, [Mg/Fe], [C/Fe], [N/Fe], and [Ca/Fe] 
as above, but now in bins of $M_K$ (mag; a proxy for 
stellar mass) rather than stellar 
velocity dispersion. Open symbols indicate cases where the primary 
measurement fell off the stellar population grid and we have 
used the median of the 
distribution of boot-strap stacks instead. The inversion program 
failed to converge for the [N/Fe] and [Ca/Fe] measurements 
for the outermost high-$M_K$ $R_e$-scaled point, so no measurement 
is shown. 
As indicated in the key, the bins have $M_K>-24.5$ (squares), 
$-24.5 > M_K > -25.5$ (pentagons), $-25.5 > M_K > -25.8$ 
(triangles), and $M_K < -25.8$ mag (circles) respectively.
We fit the radial gradients with a power law of the form 
$X = {\rm A~log} (R/R_3) + {\rm B}$ for each stellar population 
parameter $X$, where $R_3$ is either $3-6$ kpc or $1-1.5 R_e$. 
Fits to the most massive (long-dashed lines) and 
least massive (dotted lines) galaxies are shown here; all fitted coefficients 
are in Tables 1 \& 2. 
\label{fig:radialmass}}
\end{figure*}

There is now a preponderance of evidence that stellar population
properties correlate most strongly with
\sigmastar\ at their centers \citep{benderetal1993,trageretal2000a,
  gravesetal2009,wakeetal2012}.  However, there is little work
comparing radial trends as a function of mass and \sigmastar\ 
\citep{spolaoretal2010}.  We
therefore create four bins of $K-$band magnitude, based on 2MASS
photometry \citep{jarrettetal2003,skrutskieetal2006}: $M_K>-24.5$ (21
galaxies), $-24.5 > M_K > -25.5$ (23), $-25.5 > M_K > -25.8$ (24), and
$M_K < -25.8$ mag (19; Figure \ref{fig:radialmass}). We note that
there is a wide range of stellar velocity dispersion at each mass bin
(Figure \ref{fig:FJ}), so this binning scheme is truly different from
the dispersion bins presented above.

As above, we fit each stellar population property as a function of
radius, tabulated in Tables 1 \& 2. Each effective stellar population
parameter is shown as a function of $M_K$ in Figure
\ref{fig:sigmasscor} (right). Starting as in \S 5.1 by focusing on the
central measurements, we see that in general, there is more dispersion
in stellar population properties (particularly metallicity and
[Mg/Fe]) in a given $M_K$ bin as compared with a given
\sigmastar\ bin. In particular, the second $M_K$ bin ($-24.5$ to
$-25.5$ mag) shows considerably larger variance than the other
bins. We believe that this larger error bar reflects a genuine
increase in the spread in stellar populations in this bin. At yet
lower $M_K$, we no longer have a representative sample of objects. We
also generally see weaker trends in the stellar populations with $M_K$
than with \sigmastar, particularly in physically scaled bins (Table
3). The exception is [C/Fe], which shows a more significant
correlation with $M_K$ ($12 \, \sigma$). We see no correlation with
age or [\alp/Fe].  There is, however, a marginal correlation with
\feh\ ($3 \, \sigma$) and a corresponding weak correlation with [Z/H]
(also $3 \,\sigma$). If we instead consider measurements from
$1.5R_e$, we find no significant trends between stellar population
properties and $M_K$ (not even stellar age).

\begin{figure*}
\vbox{ 
\vskip -1mm
\hskip -5mm
\psfig{file=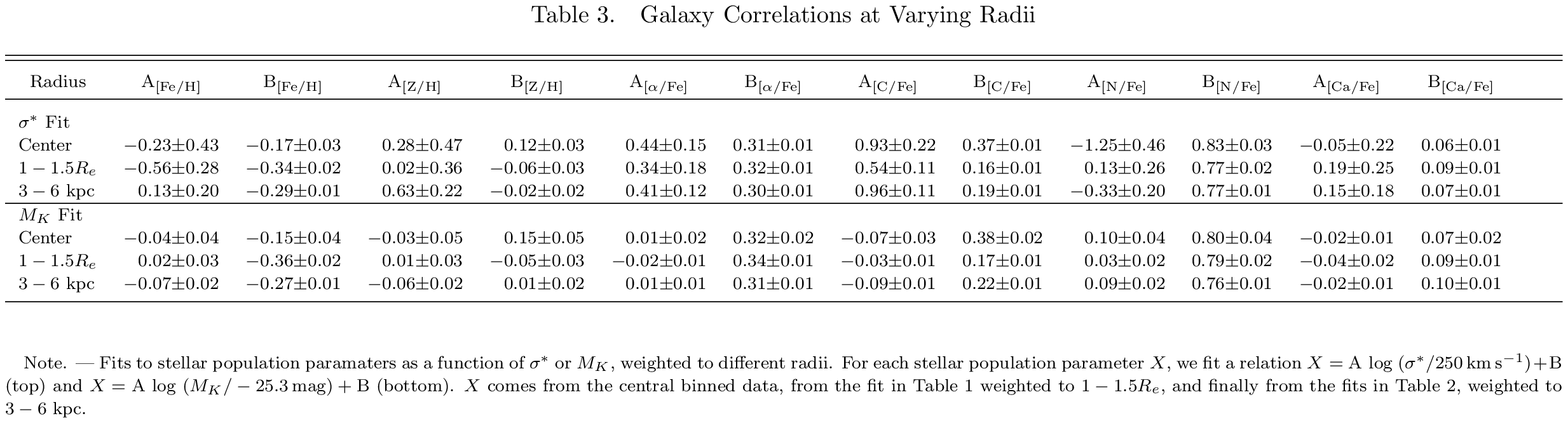,width=0.27\textwidth,keepaspectratio=true,angle=-90}
}
\vskip -0mm
\end{figure*}

\subsubsection{Bins in $M^*$ at fixed \sigmastar}

\begin{figure*}
\vbox{ 
\vskip -10mm
\hskip +10mm
\psfig{file=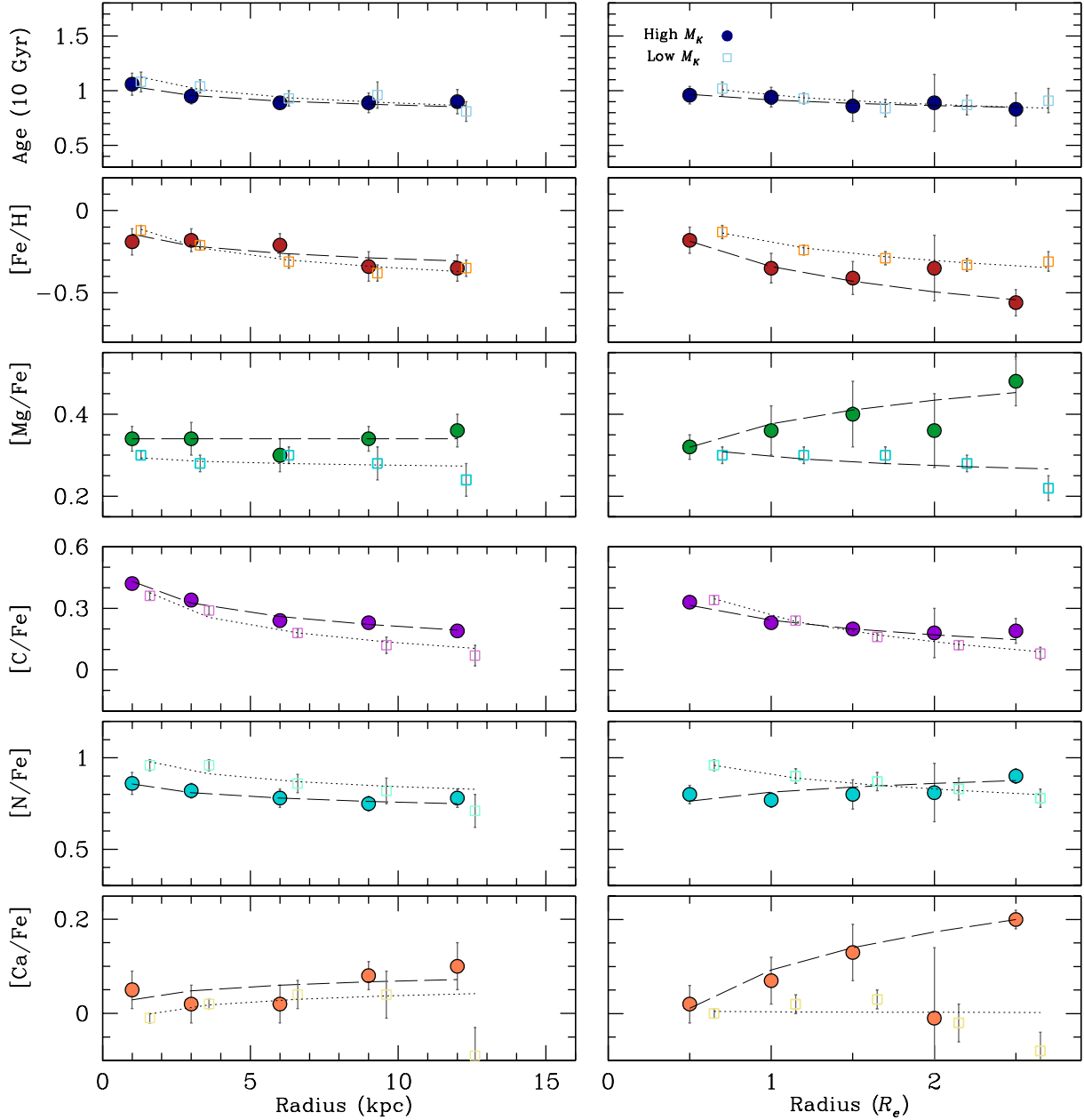,width=0.9\textwidth,keepaspectratio=true,angle=-90}
}
\vskip -0mm
\figcaption[]{
Radial gradients in age, \feh, [Mg/Fe], [C/Fe], [N/Fe], and [Ca/Fe] 
as above, but now in \sigmastar-matched bins of $M_K$
(mag; a proxy for stellar mass). 
Filled symbols have $M_K < -25$ mag while open symbols 
have $M_K > -25$ mag. 
We fit the radial gradients with a power law of the form 
$X = {\rm A~log} (R/R_3) + {\rm B}$ for each stellar population 
parameter $X$, where $R_3$ is either $3-6$ kpc or $1-1.5 R_e$. 
The fits to the high-mass (long-dashed lines) and 
low-mass (dotted lines) bins are shown here and in Tables 1 \& 2.
\label{fig:radialmassig}}
\end{figure*}

We now ask whether, at fixed \sigmastar, there
are residual trends in the stellar population gradients as a function
of stellar mass (Figure \ref{fig:radialmassig}).  We create two
luminosity bins divided at $M_K=-25$ mag (which approximately divides
the sample in two). We then enforce a matched distribution in
\sigmastar\ between 200-330 \kms\ by creating 100 stacks for each mass
bin, each having an identical distribution of \sigmastar\ and each
including a total of 30 galaxies.

We find small but detectable differences in the two mass bins.
Specifically, while the average stellar population ages are both $\sim
9.5$ Gyr, we find that the high-mass bin is more \alp-enhanced by
$\sim 0.1$ dex with (possibly) correspondingly lower \feh\ at fixed
$R_e$ (Table 1).  The C and N abundances and radial profiles are
consistent with each other, but there is a hint that the more massive
galaxies are also more Ca-enhanced at fixed \sigmastar.  All of these
trends could be qualitatively explained if the timescale for star
formation were shorter in the more massive systems, such that [Fe/H] 
is lower while [\alp/Fe] is higher and [Ca/H] follows [Fe/H]. 

The other possible explanation is that at fixed \sigmastar, more
massive galaxies are physically larger. This would explain why at a
fixed fraction of $R_e$, the more massive galaxies have lower
\feh. \citet{gravesetal2009} find that galaxies with the lowest
central surface brightness also have the oldest ages and lowest
\feh\ values, a similar trend to that seen here, albeit only for
galaxy centers. When we have robust size measurements for our sample
galaxies we will revisit the question of stellar populations through
the Fundamental Plane, also including radial stellar population
information.

\subsection{Interpreting Stellar Populations Using Radial Trends}

There are a number of interesting trends seen in the stellar
populations of elliptical galaxy centers that challenge our
understanding, particularly when compared with Galactic trends.
It is our hope that adding radial information will shed new light 
not only on the assembly history of ellipticals, but also on  
the nucleosynthetic yields that lead to these observed trends. 

There are a few important caveats to keep in mind as 
we interpret the observations. First, our stellar ages, particularly 
at the center, are subject to large ($\sim 2-3$ Gyr) uncertainties. 
Thus, we caution against over-interpreting the radial trends in 
age at present. Some work covering larger dynamic range in radius 
do detect clear age gradients \citep[e.g.,][]{labarberaetal2012}, 
but we do not go to large enough radius to detect the 
very low metallicity true halo component that has been seen in a 
few nearby cases \citep[e.g.,][]{harrisetal2007,williamsetal2015}. 
Second, we remain cautious about the nitrogen measurements given 
the possible flux calibration difficulties at the blue end 
of the spectrum. Third, recall that we are examining average 
trends in stacked spectra. Undoubtably there are interesting 
exceptions to all of these trends, and we plan to study the full 
range of parameters in future work.

We revisit our expectations for stellar population trends as a
function of radius in light of the recent picture that galaxies form
in two ``phases'', an initial burst of in-situ star formation creating
the central component, followed by late-time accretion of smaller,
fluffier units at larger radius \citep[e.g.,][]{naabetal2009}. For
galaxies in our mass range, the typical sizes of the central
components are measured to be 2-5 kpc at $z \approx 2$
\citep[e.g.,][]{vanderweletal2014}. When we look at the stellar
populations at this typical radius (Figure \ref{fig:sigmasscor}), we
find that age and [\alp/Fe] depend on \sigmastar\ rather than
$M_K$. Since we expect galaxies that form early to be denser and have
higher \sigmastar, a stronger correlation between age and star
formation timescale with \sigmastar\ seems natural. In contrast, 
\feh, [C/Fe], and [N/Fe] depend more on the total stellar mass 
of the system.

In addition to considering a fixed physical radius, we look at stellar
population trends at $\sim 1.5 R_e$ (Figure
\ref{fig:sigmasscor}). According to simulations, more massive galaxies
are increasingly dominated by accreted stars at radii beyond $\sim 5$
kpc \citep[e.g.,][]{oseretal2010}. Therefore we might expect the
correlation between \sigmastar\ and stellar population properties to
decrease when taken over the bulk of the stellar population. This is
what we observe. Beyond $\sim R_e$ we find no strong trend between
\sigmastar\ or $M_K$ and abundances or abundance ratios. This result
builds on what we saw in \citet{greeneetal2013}. There, we emphasized
that the stellar populations beyond $R_e$ in massive galaxies tend to
have low \feh$\sim -0.5$ dex and high [\alp/Fe]$\sim 0.3$ dex, stellar
populations that are not seen in the centers of any galaxies today
\citep[see also][]{benderetal2015}.  Here we see that beyond $R_e$,
galaxies over a relatively wide range in \sigmastar\ and $M_K$ have
similar abundances and abundance ratios, as expected if the more
massive galaxies were built by accreting the less massive. Only
stellar age is still seen to increase at higher \sigmastar\ when
examined at $\sim R_e$; we await better age measurements to verify
this result.

Another ongoing discussion in the literature regards carbon.  As is
seen here, [C/Fe] is observed to increase with \sigmastar\ in
elliptical galaxy centers
\citep[e.g.,][]{trageretal1998,gravesetal2007,johanssonetal2012,
  conroyetal2014}. This is in contrast to the behavior of carbon in
the Milky Way or Local Group dwarf galaxies
\citep[e.g.,][]{kirbyetal2015}.  To get such super-solar carbon levels
in the short timescales implied by the high ratio of [\alp/Fe], carbon
must come not only from intermediate-mass (AGB) stars but also from
massive stars \citep{gravesetal2007,tangetal2014}.  Carbon yields from
massive stars are thought to increase with increasing metallicity due
to increased mass-loss from winds \citep[e.g.,][]{maeder1992}.  These
same high yields at high metallicity are also invoked to explain
abundance trends in Milky Way stars \citep[e.g.,][]{henryetal2000}. If
true, we would expect that as the metallicity decreases outwards in
these elliptical galaxies, the [C/Fe] would also decrease. This is
what we observe \citep{greeneetal2013}. Interestingly, we find that
the [C/Fe] gradient follows the decline in [Fe/H] in both the
\sigmastar\ and $M_K$ bins.

There is also considerable debate in the literature about the origin
and behavior of nitrogen in elliptical galaxy centers
\citep[e.g.,][]{kelsonetal2006,johanssonetal2012}.  Our surprising
results are two-fold. First, [N/Fe] is remarkably super-solar. Even
assuming solar [O/Fe] (at odds with the observed [Mg/Fe]) [N/Fe] is
found to be three times the solar value. Second, since the N is
produced by C through the CNO cycle, their different behavior with
radius is non-intuitive (and may point to a changing O abundance
as well).  As discussed above, the flux calibration at the blue end of
the spectrum, containing CN, is quite uncertain.  We do find good
agreement between CN as measured from the SDSS spectra and the central
Mitchell fiber. We also confirm that the CN measurements do not depend
on how we treat the continuum.  Including or excluding
the overall continuum level in the stacks changes the CN1 measurement
by $<0.005$ mag, resulting in very small $< 0.02$ dex in the [N/Fe]
ratio.  To really confirm these high N abundance ratios at large
radius, we would like to perform full spectral modeling to mitigate
the impacts of blending and the uncertainties introduced by oxygen.
In the meantime, it is interesting to note other stellar systems that
display very super-solar nitrogen abundance ratios.  For instance, in
globular clusters, the wide range in [N/Fe] strongly suggests
pre-enrichment by a previous early epoch of star formation
\citep[e.g.,][]{cohenetal2005}.

Finally, we come to Ca. While nominally an \alp-element, it has long
been known, based on both Ca4227 in the blue and the calcium 
triplet (CaT) index in the
red, that Ca is under-abundant with respect to the other \alp\ elements
\citep[e.g.,][]{cohen1979,vazdekisetal1997,worthey1998,
  proctorsansom2002,terlevichetal1990,peletieretal1999,sagliaetal2002,
  thomasetal2003,choietal2014}. Like \feh, [Ca/Fe] shows no dependence
on \sigmastar. There are a number of explanations in the literature
for the CaT measurements, including changes in the initial mass
function \citep[e.g.,][]{cenarroetal2004} or a minority metal-poor
population \citep{sagliaetal2002}. However, to explain both the blue
and red index behavior, it is more natural to presume that Ca behaves
like an Fe-peak element because it is predominantly produced in Type
1a supernovae \citep[e.g.,][]{worthey1998,wortheyetal2011,
conroyetal2014}. As expected in that case, we measure a flat [Ca/Fe] ratio
with radius. The one intriguing difference is the possible increase in
[Ca/Fe] at large radius, which is worth pursuing.

\subsection{Bins of Group Richness}

\begin{figure*}
\vbox{ 
\vskip -10mm
\hskip +10mm
\psfig{file=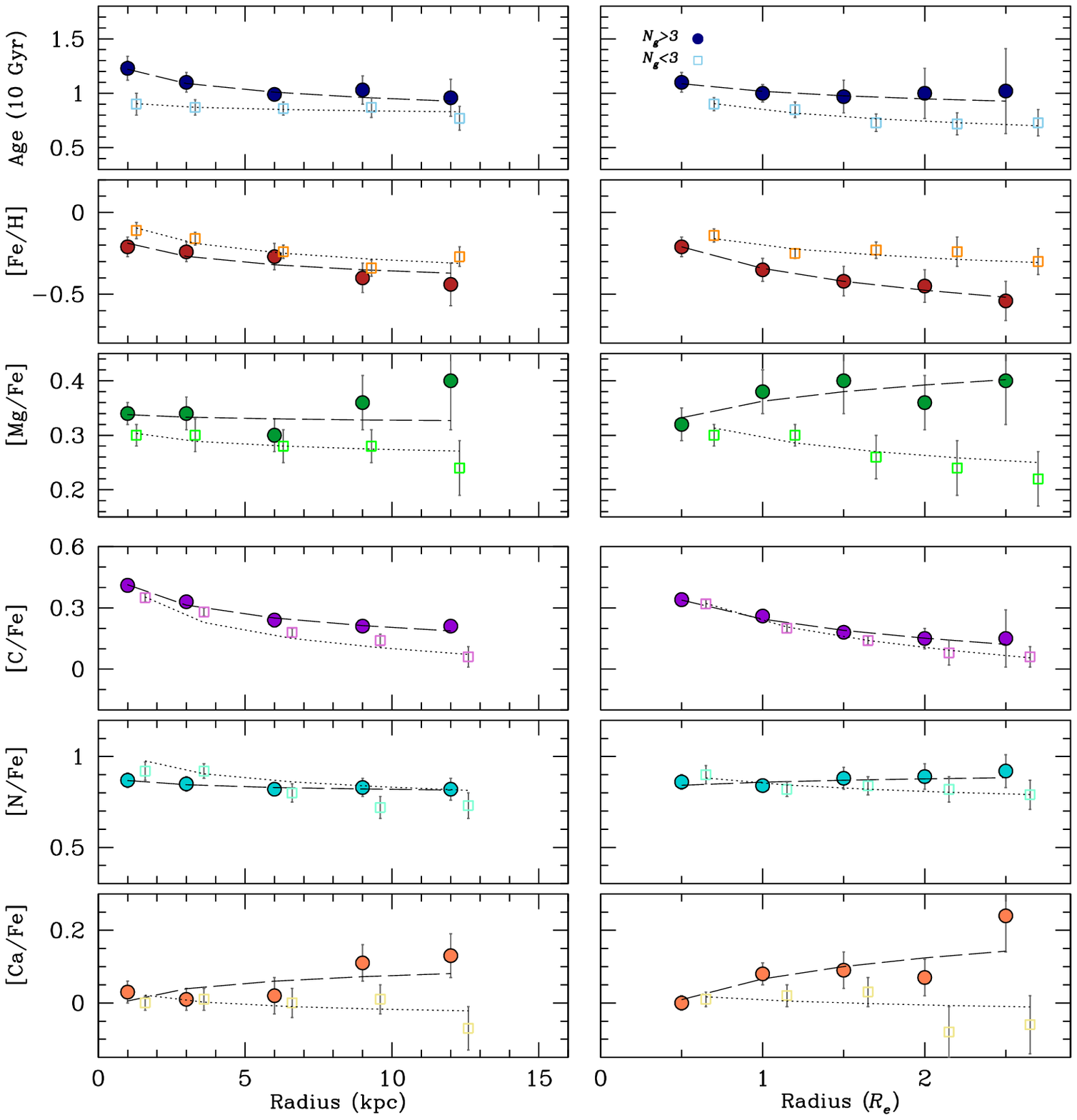,width=0.9\textwidth,keepaspectratio=true,angle=-90}
}
\vskip -0mm
\figcaption[]{Radial gradients in age, \feh, [Mg/Fe], [C/Fe], [N/Fe], and [Ca/Fe] 
as above, but now in bins of group richness, controlling for the 
distribution of \sigmastar. 
Low density (open squares) have three or fewer neighbors 
with $L > L^*$, while filled circles include everything else.
We fit the radial gradients with a power law of the form 
$X = {\rm A~log} (R/R_3) + {\rm B}$ for each stellar population 
parameter $X$, where $R_3$ is either $3-6$ kpc or $1-1.5 R_e$. 
The fits to the rich (long-dashed lines) and 
poor (dotted lines) bins are shown here and in Tables 1 \& 2.
\label{fig:radialenvmatch}}
\end{figure*}

While there are well-documented differences in the morphological mix
of galaxies as a function of local galaxy density
\citep{dressler1980}, the observations of environmental differences in
stellar population properties are quite subtle
\citep[e.g.,][]{thomasetal2005,zhuetal2010,lacknergunn2013}. Early
studies found evidence for younger ages in `field' galaxies
\citep[e.g.,][]{terlevichforbes2002}. A number of other studies
report no change in scaling relations between stellar population
parameters and \sigmastar\ as a function of local environment
\citep[e.g.,][]{kuntschneretal2002, bernardietal2006}, but do find a
larger fraction of ``rejuvenated'' galaxies with recent star formation
in low-density environments
\citep[e.g.,][]{annibalietal2007,thomasetal2010}.  Recently, the
samples have grown large enough to evaluate not just the average
properties of field and cluster galaxies, but to control for stellar
and halo mass \citep[e.g.,][]{pasqualietal2010}. For instance,
Pasquali et al.\ find that satellite galaxies at fixed mass grow older
and more metal rich as their host halo mass increases.

There is precious little literature on radial gradients in stellar
populations as a function of group richness, although a few
photometric studies find steeper metallicity gradients in
lower-density environments \citep[][]{koim2005,labarberaetal2005}. 
Given our large sample and wide range of measured group richness
\citep{maetal2014}, we are in a unique position to examine radial
stellar population trends with environment.

To divide the sample by richness, we use the group catalog of
\citet{crooketal2007}. When we divide our galaxies based
on group richness alone, the distributions in \sigmastar\ do not
match.  Instead, there are more high \sigmastar\ galaxies in richer
groups.  Since \sigmastar\ is strongly correlated with
stellar population properties, we must match the
\sigmastar\ distributions across different halo mass bins. Therefore,
we divide the sample in half based on the number of neighbors: ``low''
comprises galaxies with no more than three companions of $>L^*$ while
``high'' comprises the rest.  The raw distribution of \sigmastar\ have
median $\langle$\sigmastar$\rangle= 240$~\kms\ for the low-density bin
and $\langle$\sigmastar$\rangle= 260$~\kms\ for the high-density
bin. Since there are not sufficient numbers to further subdivide the
galaxies into bins of \sigmastar, we create 100 stacks, drawing from
the objects with \sigmastar\ between 200 and 330~\kms. We force the
low-density and high-density stacks to have the same number of objects
(24 in this case) with the same distribution of \sigmastar. We do not
have sufficient numbers to match on $M_k$ as well; two-thirds of the
rich galaxies are also in the brighter half of the sample. The results
are shown in Figure \ref{fig:radialenvmatch}.

There are only very slight differences between the two galaxy samples
divided by group richness.  The high-density stack is slightly older,
has slightly lower \feh, and slightly higher [\alp/Fe], similar to
some previous studies
\citep[e.g.,][]{bernardietal2006,clemensetal2009,cooperetal2010}. The
basic interpretation is that objects found in the highest density
peaks today likely formed earlier, and thus have older ages, higher
[\alp/Fe], and slightly lower \feh\ (although they maintain roughly
solar metallicity overall). Again, the other possibility is that
  at fixed \sigmastar, galaxies in denser environments tend to be
  slightly larger. We plan to control for galaxy size in future work.
Turning to the other light elements, we find that [C/Fe] is marginally
higher in the high-density bin, while [N/Fe] and [Ca/Fe] are
comparable between the two.  Finally, it is interesting to note that
the sample variance is larger for the low-density sample.

We see marginal evidence for a steeper \feh\ gradient in the higher
halo-mass bin as a function of effective radius (Figure
\ref{fig:radialenvmatch}, right; $-0.44 \pm 0.14$ for the higher halo
mass bin, $-0.2 \pm 0.1$ for the lower halo mass bin). The slope
difference is only significant at $\sim 2 \, \sigma$. [\alp/Fe] also
shows a very marginal difference in the opposite direction, with
slopes of $0.1 \pm 0.08$ and $-0.09 \pm 0.06$ for the higher and lower
halo mass bins respectively. Because of the anti-correlation between
\feh\ and [\alp/Fe], the resulting gradients in [Z/H] for the two bins
are comparable ($-0.3 \pm 0.1$ in both cases).  If this result is
confirmed, it perhaps suggests that at a given \sigmastar, galaxies in
richer environments are more compact, and thus show steeper
\feh\ gradients.  While we do see marginal differences between
\feh\ and [\alp/Fe], we do not confirm results from previous
photometric surveys that reported steeper metallicity gradients in
lower density environments \citep{koim2005,labarberaetal2005}.

\begin{figure*}
\vbox{ 
\vskip -15mm
\hskip +10mm
\psfig{file=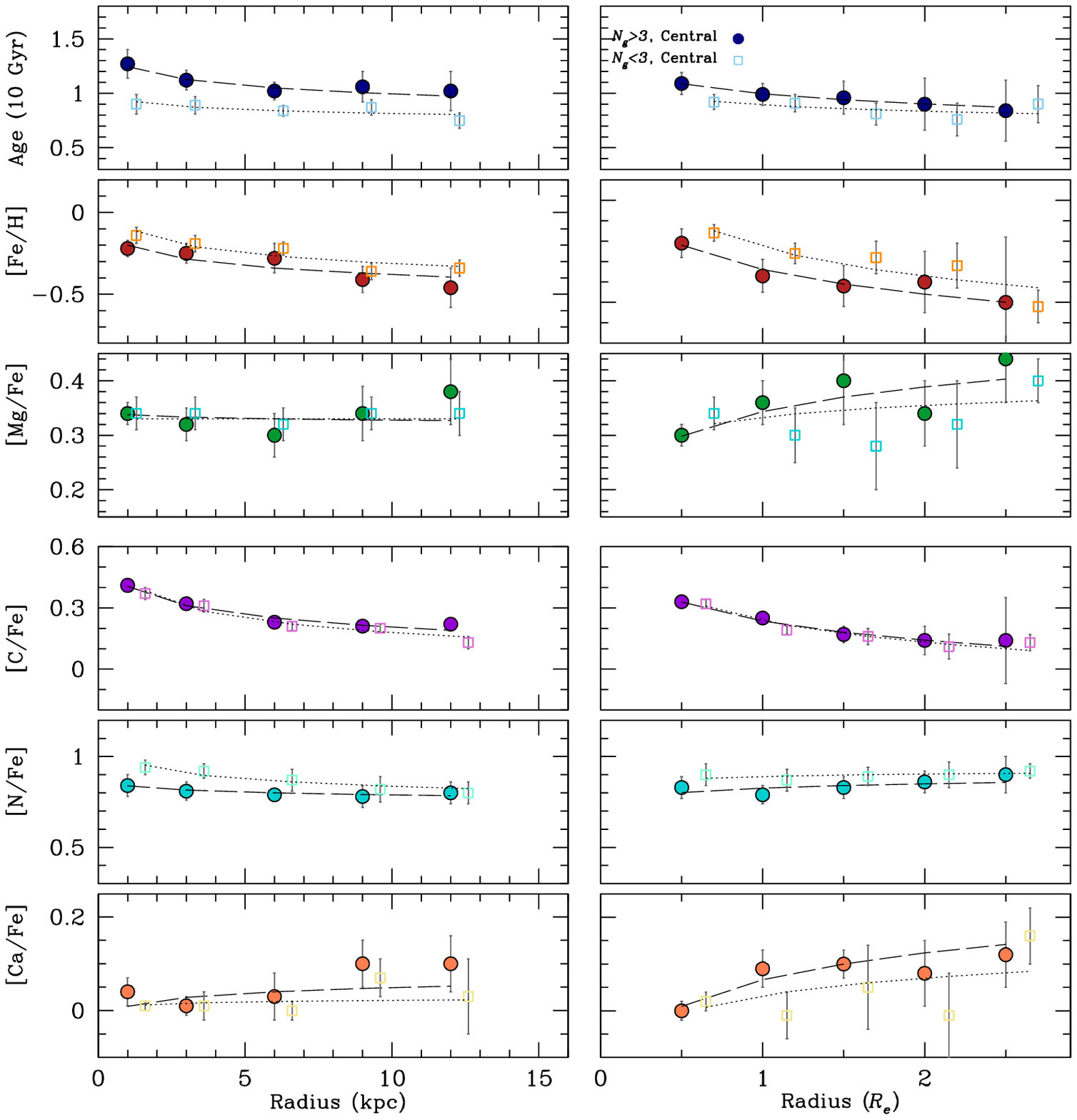,width=0.9\textwidth,keepaspectratio=true,angle=-90}
}
\vskip -0mm
\figcaption[]{
Radial gradients in age, \feh, [Mg/Fe], [C/Fe], [N/Fe], and [Ca/Fe] 
as above, but now in bins of group richness for so-called 
central galaxies (the most massive galaxy in the group, according 
the the Crook et al. 2007 catalog). As above, we divide the sample into 
those with three or fewer neighbors, and all the rest. We also control the 
distribution in \sigmastar\ to match between the two bins.
We fit the radial gradients with a power law of the form 
$X = {\rm A~log} (R/R_3) + {\rm B}$ for each stellar population 
parameter $X$, where $R_3$ is either $3-6$ kpc or $1-1.5 R_e$. 
The fits to the rich (long-dashed lines) and 
poor (dotted lines) bins are shown here and in Tables 1 \& 2.
\label{fig:radialbcg}}
\end{figure*}

\citet{labarberaetal2014} argue that in addition to looking for trends
as a function of halo mass, we should also be dividing the samples
into ``central'' (the most massive galaxy in a halo) and ``satellite''
galaxies. We therefore repeat the stacking exercise, but this time
removing all satellite galaxies (Figure \ref{fig:radialbcg}); we do
not have sufficient numbers of satellite galaxies to stack them
alone. Interestingly, when we examine central galaxies alone, the
marginal stellar population differences discussed above vanish
(although we note that our statistical power is lessened by the
smaller sample size as well).  Perhaps the differences between the two
samples are driven by differing fractions of satellite galaxies in the
two environmental stacks, but better statistics are needed before we
can be sure. On the other hand, studies of individual brightest
cluster galaxies in rich clusters find a wide spread in central
properties such as age and [\alp/Fe] \citep[e.g.,][]{olivaetal2015},
as well as evidence for distinct accretion episodes
\citep[e.g.,][]{coccatoetal2010,coccatoetal2011}.  In future work we
will investigate in more detail the spread in central galaxy
properties as a function of halo mass.

We do not reproduce the trend found by La Barbera et al.\ that the
central galaxies are younger in the larger halos. Instead, when we
focus on central galaxies only, we see no significant difference
between the stellar populations of the two groups at the 0.1 dex
level.  On the other hand, ``low-mass'' halos in the La Barbera study
have $M_h < 10^{12.5}$~\msun. This is considerably lower than
the likely halo mass of our galaxies, given that their stellar
masses reach $M^* \approx 10^{12}$~\msun.  Furthermore, the
differences reported by La Barbera et al.\ are at the 0.025 dex level,
not yet accessible with our data.

\section{Summary}
\label{sec:Summary}

Using integral-field spectroscopy, we have looked at the average
stellar population gradients for a large sample of 100 early type
massive galaxies. We are able to reach radii $\sim 2.5 R_e$ or $\sim
15$ kpc.  In keeping with previous results, we find no significant
gradients in stellar population age nor [\alp/Fe] abundance ratios
with radius, and gentle gradients in \feh. We thus confirm our
previous result that the stellar populations in the outskirts of
massive galaxies have sub-solar \feh\ but are enhanced in [\alp/Fe],
suggesting that the stars formed quickly and early, but in shallow
potentials \citep{greeneetal2013,benderetal2015}.

We examine the stellar population properties weighted towards 3-6 kpc,
the typical sizes of massive galaxy cores as observed at $z \approx 2$
\citep[e.g.,][]{vanderweletal2014}. We see that age and [\alp/Fe] rise
with increasing \sigmastar, as we might expect if denser galaxies with
higher \sigmastar\ form earlier. We also find that at fixed physical
radius, \feh\ and [C/Fe] correlate more strongly with $M_K$ than
stellar velocity dispersion.  In contrast, when looking at bins
weighted towards $\sim R_e$, we find no strong trends between
abundances or abundance ratios and \sigmastar\ or $M_K$. The average
star as measured near the half-light radius in the most massive
ellipticals is similar to the average star as measured at the
half-light radius in galaxies of lower mass, as we might expect if
large galaxies grow via accreting smaller satellites.  We do, however,
still see a trend between stellar age and \sigmastar\ even in the
$R_e$-weighted bins.

The gradients in [C/Fe] are similar to those seen in \feh. We suggest
that the C comes mainly from mass loss in massive stars because there
is not time to get it from intermediate-mass AGB stars
\citep[e.g.,][]{gravesetal2007}. Higher yields due to mass loss from
metal-rich stars \citep[e.g.,][]{maeder1992} cause a pseudo-secondary
dependence of carbon on Fe.  In contrast, we see super-solar [N/Fe]
that persists to large radius; the mismatch between [C/Fe] and [N/Fe]
is a puzzle.  [Ca/Fe] has solar values over the entire radial range
that we observe (with the possible exception of a rise at large radius
in the high dispersion bin), consistent with the idea that significant
Ca is produced in Type 1a supernovae.

Thanks to our relatively large sample, we are able to examine trends
in stellar mass at fixed \sigmastar.  At fixed \sigmastar, we find
marginal evidence that galaxies with higher stellar mass are more
\alp-enhanced and [Ca/Fe] enhanced (because they are slightly
\feh\ poor), suggesting a shorter timescale for star formation in more
massive systems.

Finally, we perform one of the most extensive spectroscopic studies 
of stellar population gradients as a function of group richness, 
while controlling for \sigmastar. Overall, the differences in stellar 
population properties at large radius as a function of richness are 
very small, suggesting that internal properties like \sigmastar\ 
determine stellar population gradients 
\citep[as in galaxy centers; e.g.,][]{zhuetal2010}.
Galaxies in richer environments ($N_{\rm
  nei} > 3$) tend to be slightly older, slightly \alp-enhanced, and
slightly \feh\ poor. We also see very slight trends towards shallower 
declines in \feh\ in lower-mass halos. When we
restrict attention to only central galaxies, these slight differences
vanish, perhaps suggesting that they are driven by the fraction of
satellite galaxies in the stacks.  These trends are quite weak. 
Better statistics are needed to confirm them. Furthermore, it will 
be quite interesting to combine our dynamical and stellar population 
information \citep[e.g.,][]{raskuttietal2014,jimmyetal2013,naabetal2014}.

By the end of the MASSIVE survey, we should roughly double the number
of $M_K < -25.3$ mag galaxies in the sample, improving our
ability to examine trends with environment and mass at fixed
\sigmastar. Our sample will be further complemented by ongoing 
ambitious integral-field galaxy surveys such as 
CALIFA \citep{sanchezetal2012}, MaNGA \citep{bundyetal2015}, 
and SAMI \citep{croometal2012}.
When combined with our dynamical constraints on the total
masses of these galaxies, these stellar population constraints on the
age and the mass-to-light ratios of the galaxies will help address a
number of pressing questions in galaxy evolution, including the
dependence of the initial mass function of \sigmastar, the ratio of
black hole mass to stellar mass at the high mass end, and the role 
of dark matter halo mass in the internal evolution of 
massive galaxies.

\acknowledgements

We thank the referee for a prompt and helpful report. 
C.P. Ma and J. E. Greene acknowledge funding from NSF grants
AST-1411945 and AST- 1411642.  NJM is supported by the Beatrice Watson
Parrent Fellowship. JEG gratefully acknowledges conversations with J.E. 
Gunn and R. Schiavon. This research has made use of the NASA/IPAC
Extragalactic Database (NED) which is operated by the Jet Propulsion
Laboratory, California Institute of Technology, under contract with
the National Aeronautics and Space Administration.
The Pan-STARRS1 Surveys (PS1) have been made possible through
contributions of the Institute for Astronomy, the University of
Hawaii, the Pan-STARRS Project Office, the Max-Planck Society and its
participating institutes, the Max Planck Institute for Astronomy,
Heidelberg and the Max Planck Institute for Extraterrestrial Physics,
Garching, The Johns Hopkins University, Durham University, the
University of Edinburgh, Queens University Belfast, the
Harvard-Smithsonian Center for Astrophysics, the Las Cumbres
Observatory Global Telescope Network Incorporated, the National
Central University of Taiwan, the Space Telescope Science Institute,
the National Aeronautics and Space Administration under Grant
No. NNX08AR22G issued through the Planetary Science Division of the
NASA Science Mission Directorate, the National Science Foundation
under Grant No. AST-1238877, the University of Maryland, and Eotvos
Lorand University (ELTE).  We thank the PS1 Builders and PS1
operations staff for construction and operation of the PS1 system and
access to the data products provided.


\end{document}